\documentclass[aps,twocolumn]{revtex4-1}
\usepackage[T1]{fontenc}
\usepackage[latin9]{inputenc}
\usepackage{amsthm}
\usepackage{amsmath}
\usepackage{graphicx}

\begin{document}

\title{Flow correlations from a hydrodynamics model with dynamical freeze-out and initial conditions based on perturbative QCD and saturation}

\author{H.~Hirvonen${}^{a,b}$, K.~J.~Eskola${}^{a,b}$, H.~Niemi${}^{a,b}$,  }
\affiliation{$^{a}$University of Jyv\"askyl\"a, Department of Physics, P.O. Box 35, FI-40014 University of Jyv\"askyl\"a, Finland}
\affiliation{$^b$Helsinki Institute of Physics, P.O.Box 64, FI-00014 University of
Helsinki, Finland}

\begin{abstract}
We extend the applicability of the hydrodynamics, perturbative QCD and saturation -based EKRT (Eskola-Kajantie-Ruuskanen-Tuominen) framework for ultrarelativistic heavy-ion collisions to peripheral collisions by introducing dynamical freeze-out conditions. As a new ingredient compared to the previous EKRT computations we also introduce a non-zero bulk viscosity. We compute various hadronic observables and flow correlations, including normalized symmetric cumulants, mixed harmonic cumulants and flow-transverse momentum correlations, and compare them against measurements from the BNL Relativistic Heavy Ion Collider (RHIC) and the CERN Large Hadron Collider (LHC). We demonstrate that the inclusion of the dynamical freeze-out and bulk viscosity allows a better description of the measured flow coefficients in peripheral collisions and enables the use of an extended centrality range when constraining the properties of QCD matter in the future.

\end{abstract} 
 
 
\maketitle 

\section{Introduction}

Heavy-ion collisions at ultrarelativistic energies provide the means to produce and investigate experimentally quark-gluon plasma (QGP), a strongly interacting fluid of quarks and gluons.  In recent years the two main collider experiments that have investigated QGP properties are the Relativistic Heavy Ion Collider (RHIC) at Brookhaven National Laboratory (BNL), and Large Hadron Collider (LHC) at CERN. In these experiments a small, short-lived, fluid-like behaving droplet of strongly interacting matter is created at nearly zero net-baryon density. The matter properties of QGP such as its equation of state (EoS) and transport coefficients are reflected in the detailed behavior of various experimental observables; see, e.g., ~Refs.~\cite{Luzum:2008cw,Bozek:2009dw,Song:2011qa,Niemi:2011ix,Niemi:2012ry,Ryu:2015vwa,Karpenko:2015xea}. 

The equation of state of strongly interacting matter at zero net-baryon density is currently well known from lattice-QCD computations, and the expected transition temperature $T_c \approx 150-160$ MeV~\cite{Bazavov:2014pvz, Bazavov:2017dsy, Borsanyi:2013bia, Borsanyi:2010cj} from hadronic matter to QGP is well within the reach of the LHC and RHIC experiments. Currently there are some experimental constrains on the equation of state~\cite{Pratt:2015zsa, Moreland:2015dvc,Alba:2017hhe}, but even the lattice-QCD data allows some freedom in the EoS parametrizations~\cite{Auvinen:2020mpc}. The best knowledge about the transport properties of QCD matter is coming from the global fits of fluid dynamical computations to the available low-$p_T$ data from RHIC and LHC~\cite{Bernhard:2016tnd,Bass:2017zyn, Bernhard:2019bmu, JETSCAPE:2020mzn, Nijs:2020roc, Auvinen:2020mpc, Parkkila:2021tqq, Parkkila:2021yha}. Currently, at least within the given models, the shear viscosity at temperatures near the QCD transition temperature is quite well constrained. However, the same cannot be said about the bulk viscosity. Even if the different analyses are based on very similar underlying models, the final constraints on the bulk viscosity can differ quite significantly depending on the details of the selected data and fine details of the models. 

The experimental information about the collective dynamics and the spatial structure of the initial conditions is primarily encoded in the flow measurements. The most basic quantities are the Fourier components of the azimuthal hadron spectra, usually called the flow coefficients $v_n$. The measured flow coefficients reflect the collective fluid dynamical behavior of the system, as they are generated during the evolution of the system when the initial spatial inhomogeneities are converted into momentum-space anisotropies. In the fluid dynamical limit the driving force for this conversion is the inhomogeneus pressure gradients, and the effectiveness of the conversion is dictated by the EoS and the transport properties of QCD matter. 

In the actual collisions the flow coefficients fluctuate strongly from event to event, and the fluctuations need to be explicitly considered when modeling the collisions. The presence of the flow fluctuations complicates the modeling, but at the same time they offer also a possibility to probe the initial conditions and the space-time evolution in much greater detail. For example, the relative fluctuation spectra of the elliptic flow coefficient $v_2$ are practically independent of the QCD matter properties, and reflect mainly the initial density fluctuations, giving thus a way to directly constrain the initial particle production~\cite{Niemi:2012aj} at least at the LHC energies, see the discussion in Ref.~\cite{Barbosa:2021ccw}. Moreover, the various observables measuring the correlations between the flow coefficients react to the matter properties and initial conditions in a nontrivial way, and offer further constraints on both of them. In particular, the correlations cannot be trivially reproduced just by reproducing the flow coefficients themselves~\cite{Niemi:2015qia}.

The aim of this paper is to calculate various measurable flow-correlators by using relativistic second-order fluid dynamics with QCD-based initial conditions. The main ingredients that go into the computation are the matter properties, equation of state and transport coefficients, initial conditions for the fluid dynamical evolution given by the primary production of particles, and finally the conditions when the fluid dynamical evolution ceases and the fluid decouples into free hadrons. 

The initial conditions are computed by using the perturbative QCD based Eskola-Kajantie-Ruuskanen-Tuominen (EKRT) saturation model~\cite{Paatelainen:2013eea, Niemi:2015qia}, where the primary quantity is the minijet transverse energy computed in next-to-leading order perturbative QCD. The low-$p_T$ production of the particles is then controlled by a saturation conjecture, detailed in Sec.~\ref{sec:ini}. The EKRT saturation model is the main feature that gives a predictive power to our computation. Once the framework is fixed at some collision system, e.g.\ in central Pb+Pb collisions at the LHC, the collision energy, centrality, and nuclear mass number dependence of hadronic observables are predictions of the model~\cite{Niemi:2015qia, Niemi:2015voa, Eskola:2017bup, Eskola:2017imo}

Once the initial conditions are given, the remaining inputs to the fluid dynamical computation are the matter properties. The EoS is provided by the s95p parametrization of lattice-QCD results~\cite{Huovinen:2009yb}, and the specific shear viscosity $\eta/s$, is parametrized such that it has a minimum around the QCD transition temperature. As a new ingredient compared to the previous EKRT computations we introduce nonzero bulk viscosity, parametrized such that it is peaked close to $T_c$. The main impact of bulk viscosity is to reduce the average $p_T$ of hadrons~\cite{Ryu:2015vwa}. This allows us to relax our earlier~\cite{Niemi:2015qia, Niemi:2015voa, Eskola:2017bup, Eskola:2017imo} rather high chemical freeze-out temperature $T_{\rm chem} = 175$ MeV, in order to better reproduce the measured identified hadron abundances, while still reproducing the measured average transverse momentum of hadrons. 

Another new feature in the computation is the dynamical condition to decouple the system into free hadrons. The earlier EKRT results were computed using a constant-temperature decoupling at $T_{\rm dec} = 100$ MeV. It can be argued that the system decouples when the mean free path of hadrons is larger than the size of the system. The mean free path is a function of temperature, and if the system size is fixed the condition gives a constant temperature. However, the system size actually changes as function of time when the system expands, and moreover the system size varies from collision to collision: Central nuclear collisions produce a much larger system than peripheral ones. In order to account for the differences in the size of the systems, we introduce two conditions for decoupling. The global condition compares the overall size of the system to the mean free path, or here rather to the relaxation time in the second-order fluid dynamics, and the local condition that requires that the Knudsen number $\rm Kn$, the ratio of microscopic and macroscopic length or time scales, is sufficiently small the fluid dynamics to be applicable~\cite{Huovinen:2008te, Gallmeister:2018mcn}. We note that this approach, in particular the global condition, is slightly different from the earlier works where dynamical decoupling was developed~\cite{Eskola:2007zc, Ahmad:2016ods}.

The main advantage of using dynamical decoupling, besides that it is physically better motivated than the constant-temperature decoupling, is that it allows one to extend the agreement between the fluid computation and the measured flow coefficients towards peripheral nuclear collisions. In particular, the success of fluid dynamics in reproducing the flow coefficients in high-multiplicity proton-nucleus collisions \cite{Bozek:2012gr, Bozek:2013uha, Bozek:2013ska, Qin:2013bha, Werner:2013ipa, Kozlov:2014fqa,Romatschke:2015gxa, Shen:2016zpp} suggests that fluid dynamical models should then also describe peripheral nuclear collisions with similar hadron multiplicities.

This paper is organized in the following way: In Sec.~\ref{sec:ini} we shortly review the EKRT saturation model. In Sec.~\ref{sec:fluidsetup} we introduce the second-order fluid dynamics, and give the parametrizations of shear and bulk viscosities, and the corresponding corrections to the hadron momentum distributions. In Sec.~\ref{sec:freeze-out} we detail the dynamical freeze-out conditions, and in Sec.~\ref{sec:correlators} we introduce the definitions of the experimental observables. The results from the computations are given in Sec.~\ref{sec:results}, where we show the new results with bulk viscosity and dynamical decoupling and compare those to the earlier predictions of the EKRT model. Finally the summary and conclusions are given in Sec.~\ref{sec:conclusions}.

\section{Initial conditions}
\label{sec:ini}

The initial energy density profile is computed by using the EKRT saturation model~\cite{Eskola:1999fc, Paatelainen:2012at, Paatelainen:2013eea, Niemi:2015qia}. It is based on the next-to-leading-order perturbative QCD (pQCD) computation of transverse energy ($E_T$) production, controlled by the low-$p_T$ cutoff scale $p_0$ determined from the local saturation condition \cite{Paatelainen:2012at},
\begin{equation}
\frac{{\rm d}E_T}{{\rm d}^2{\bf r}}(T_A T_A (\mathbf{r}), p_0,\sqrt{s_{\rm NN}}, A, \Delta y, {\bf b}, \beta) = \left (\frac{K_{\rm{sat}}}{\pi} \right ) p_0^3 \Delta y,
\label{eq: saturation}
\end{equation}
where $\Delta y$ is the rapidity interval, $\bf b$ is the impact parameter, $K_{\rm sat}$ quantifies the uncertainty in the onset of saturation, and $\beta$ quantifies the freedom in the NLO $E_T$ definition with low-$p_T$ cutoff. The solution $p_0 = p_{\rm sat}$ of the saturation condition then inherits the $\sqrt{s_{\rm NN}}$ and $A$ dependence from the NLO pQCD computation of $E_T$, and the nuclear geometry enters through the product $T_A T_A$ of the nuclear thickness functions,
\begin{equation}
p_{\rm sat} = p_{\rm sat}\left( T_A T_A (\mathbf{r}), \sqrt{s_{\rm NN}}, A, \Delta y, {\mathbf b}, K_{\rm sat}, \beta\right).
\end{equation}
The local energy density at the formation time $\tau_s = 1/p_{\rm sat}$ can then be written using $p_{\rm sat}$ as 
\begin{equation}
e(\mathbf{r}, \tau_s(\mathbf{r}) ) = \frac{dE_T(p_{\rm sat})}{d^2{\bf r}} \frac{1}{\tau_s(\mathbf{r})\Delta y} = \frac{K_{\rm{sat}}}{\pi}[p_{\rm sat}(\mathbf{r})]^4.
\label{eq:edensity}
\end{equation}
At each point in the transverse plane the energy density is further evolved into a common initialization time $\tau_0 = 1/p_{\rm sat, min} \approx 0.2$ fm by using $(0+1)$-dimensional Bjorken expansion, where the minimum saturation scale $p_{\rm sat, min} = 1$ GeV. Below this scale the computed energy density profile is connected smoothly to the $e \propto T_A T_A$ profile. As in the earlier works, we take $\beta = 0.8$, and $K_{\rm sat}$ is fixed from the charged 
particle multiplicity measured in central $\sqrt{s_{\rm NN}} = 2.76$ TeV Pb+Pb collisions. For further details and explicit parametrizations of $p_{\rm sat}$, see Refs.~\cite{Niemi:2015qia, Niemi:2015voa, Eskola:2017bup}. 

The nuclear thickness functions are computed by first randomly sampling the nucleon positions from the Woods-Saxon nucleon density profiles. The Au and Pb nuclei are taken as spherical with a radius $R = 6.38 (6.7)$ fm for Au (Pb), and a thickness parameter $d=0.55$ fm. As in Ref.~\cite{Giacalone:2017dud}, in the case of Xe we take into account the deformation by introducing the parameters $\beta_2 = 0.162$ and $\beta_4 = -0.003$~\cite{Moller:2015fba}. The Xe radius is $R = 5.49$ fm and the thickness parameter $d = 0.54$ fm. 

The nuclear thickness functions are then computed by summing up the individual nucleon thickness functions,
\begin{equation}
T_A(\mathbf{r}) = \sum_i T_{n,i} (\mathbf{r}_i - \mathbf{r}), 
\end{equation}
where $T_n$ is a Gaussian with a width $\sigma = 0.43$ fm. The event-by-event fluctuations emerge from the random positions of the nuclei, and impact parameter: The fluctuating $T_A T_A$ profile leads to a fluctuating energy density profile through the $T_A T_A$-dependence of the saturation scale in Eq.~(\ref{eq:edensity}). 

A randomly sampled collision event, i.e.\ the nucleon positions in the nuclei and the impact parameter between the two nuclei, is accepted using a geometric criterion: We require that there is at least one pair of colliding nucleons with a transverse distance less than $\sqrt{\sigma_{\rm NN}/\pi}$, where $\sigma_{\rm NN}$ is the inelastic nucleon-nucleon cross section. Here we take $\sigma_{\rm NN} = 42$ mb in $\sqrt{s_{\rm NN}} = 200$ GeV Au+Au, $\sigma_{\rm NN} = 64$ mb in $\sqrt{s_{\rm NN}} = 2.76$ TeV Pb+Pb, $\sigma_{\rm NN} = 70$ mb in $\sqrt{s_{\rm NN}} = 5.023$ TeV Pb+Pb, and $\sigma_{\rm NN} = 72$ mb in $\sqrt{s_{\rm NN}} = 5.44$ TeV Xe+Xe collisions. We emphasize that this criterion is only used as a condition that nuclear collision happens at all, it is not needed in the computation of the initial profile.

\section{Fluid dynamical evolution and particle spectra}
\label{sec:fluidsetup}

After the hot strongly interacting system is produced at $\tau_0 \sim 1/p_{\rm sat}$, the subsequent spacetime evolution is computed using relativistic dissipative fluid dynamics. The basic equations of fluid dynamics are the local conservation laws of energy, momentum and conserved charges like net-baryon number. These can be expressed in terms of the energy-momentum tensor and charge 4-currents as $\partial_\mu T^{\mu\nu} = 0$, and $\partial_\mu N_i^{\mu} = 0$. In what follows we shall neglect the conserved charges so that it is sufficient to consider only the energy-momentum tensor. It can be decomposed with respect to the fluid 4-velocity $u^{\mu}$ as
\begin{equation}
T^{\mu\nu} = e u^\mu u^\nu - P \Delta^{\mu\nu} + \pi^{\mu\nu}, 
\label{eq:energymomentum}
\end{equation}
where the fluid velocity is defined in the Landau picture, i.e.\ as a time-like, normalized eigenvector of the energy momentum tensor, $T^\mu_\nu u^\nu = e u^\mu$. Here $e = T^{\mu\nu} u_\mu u_\nu$ is the local energy density, $P = -\frac{1}{3}\Delta_{\mu\nu}T^{\mu\nu}$ is the isotropic pressure, and $\pi^{\mu\nu} = T^{\langle \mu\nu \rangle}$ is the shear-stress tensor. The angular brackets denote the projection operator that takes the 
symmetric and traceless part of the tensor that is orthogonal to the fluid velocity, i.e., $A^{\langle\mu\rangle} = \Delta^{\mu\nu}A_\nu$ and 
\begin{equation}
A^{\langle\mu\nu\rangle} = \frac{1}{2}\left[\Delta^{\mu}_{\alpha}\Delta^{\nu}_{\beta} + \Delta^{\mu}_{\beta}\Delta^{\nu}_{\alpha} - \frac{2}{3} \Delta^{\mu\nu} \Delta_{\alpha\beta}\right]  A^{\alpha\beta}, 
\end{equation} 
where $\Delta^{\mu\nu} = g^{\mu\nu} - u^\mu u^\nu$, and $g^{\mu\nu}$ is the metric tensor for which we use the $g^{\mu\nu} = \rm diag(+, -, - ,-)$ convention. The bulk viscous pressure is defined as $\Pi = P - P_0$, where $P$ is the total isotropic pressure and $P_0$ is the equilibrium pressure. 

The conservation laws are exact, but they do not give sufficient constraints to solve the evolution. The simplest fluid dynamical theory follows by neglecting the dissipative effects completely. In that case the system is always in a strict thermal equilibrium, entropy is conserved, and the equation of state in the form $P_0 = P_0(e)$ closes the system. The dissipation plays, however, a significant role in the evolution of the system in heavy-ion collisions, and it cannot be readily neglected. The dissipative effects are contained in the shear-stress tensor and in the bulk viscous pressure. Therefore the remaining task is to write evolution equations for them. In the formalism of Israel and Stewart~\cite{Israel:1979wp} the equations take the form
\begin{equation}
 \tau_\Pi \frac{d}{d\tau}\Pi+ \Pi = -\zeta \theta  - \delta_{\Pi\Pi}\Pi\theta + \lambda_{\Pi\pi}\pi^{\mu\nu}\sigma_{\mu\nu},
\label{eq:IShydrobulk}
\end{equation} 

\begin{eqnarray}
 \tau_\pi \frac{d}{d\tau}\pi^{\langle \mu \nu \rangle} + \pi^{\mu\nu}  =& 2\eta \sigma^{\mu\nu}  + 2 \tau_\pi \pi_\alpha^{\langle \mu}\omega^{\nu\rangle \alpha} \notag \\ 
 & - \delta_{\pi\pi} \pi^{\mu\nu} \theta-\tau_{\pi\pi} \pi_{\alpha} ^{\langle \mu} \sigma^{\nu\rangle \alpha} \\
 &+ \varphi_7 \pi_\alpha^{\langle \mu} \pi^{\nu\rangle \alpha}+\lambda_{\pi\Pi} \Pi \sigma^{\mu\nu} \notag,
\end{eqnarray}
where $\sigma^{\mu\nu} = \nabla^{\langle \mu}u^{\nu\rangle}$ is the strain-rate tensor, $\omega^{\mu\nu} = \frac{1}{2}\left(\nabla^{\mu}u^{\nu} - \nabla^{\nu}u^{\mu}\right)$ is the vorticity tensor, and $\theta = \nabla_{\mu}u^{\mu}$ is the expansion rate. The shear and bulk relaxation times are denoted by $\tau_\pi$ and $\tau_\Pi$ respectively, while first-order transport coefficients are the shear viscosity $\eta$ and the bulk viscosity $\zeta$. The coefficients of the nonlinear terms $\delta_{\Pi\Pi}, \lambda_{\Pi\pi}, \delta_{\pi\pi}, \tau_{\pi\pi}, \varphi_7, \lambda_{\pi\Pi}$ are second-order transport coefficients. Formally these equations can be derived from kinetic theory~\cite{Israel:1979wp, Denicol:2010xn, Denicol:2012cn, Denicol:2012es, Molnar:2013lta, Betz:2008me, Betz:2010cx, Denicol:2011fa}, by expanding around equilibrium and keeping terms up to the first order in gradients (or Knudsen number, a ratio of microscopic and macroscopic time/length scales, such as ${\rm Kn}\sim \tau_\pi \nabla_{\mu}u^{\mu}$, \cite{Niemi:2014wta}), second order in inverse Reynolds number $\sim \pi^{\mu\nu}/P_0$, and product of Knudsen number and inverse Reynolds number.

In this work the fluid dynamical setup is the same as in our previous works \cite{Niemi:2012ry, Niemi:2011ix, Niemi:2015qia, Niemi:2015voa, Eskola:2017bup}, i.e.\ we assume boost-invariant longitudinal expansion, so that it is enough to solve the equations of motion numerically in (2+1) dimensions~\cite{Molnar:2009tx}. The second-order transport coefficients in the Israel-Stewart equations are taken from the 14-moment approximation to massless gas~\cite{Denicol:2010xn, Denicol:2012cn, Molnar:2013lta} and bulk-related coefficients are from Ref.~\cite{Denicol:2014vaa}, i.e.,
\begin{equation}
    \begin{split}
        &\delta_{\Pi\Pi} = \frac{2}{3}\tau_\Pi, \hspace{2mm} \lambda_{\Pi\pi} = \frac{8}{5}\Big(\frac{1}{3}-c_s^2\Big)\tau_\Pi, \hspace{2mm} \delta_{\pi\pi} = \frac{4}{3}\tau_\pi \\
        &  \tau_{\pi\pi} = \frac{10}{7} \tau_\pi, \hspace{2mm} \varphi_7 = \frac{9}{70 P_0}, \hspace{2mm} \lambda_{\pi\Pi} = \frac{6}{5} \tau_\pi,
    \end{split}
\end{equation}
where $c_s^2$ is the speed of sound. The shear and bulk relaxation times are given by
\begin{equation}
 \tau_\pi = \frac{5\eta}{e+P_0}, \hspace{3mm} \tau_\Pi = \left(\mathrm{15} \Big(\frac{1}{3}-c_s^2\Big)^2(e+P_0)\right)^{-1}\zeta.
 \label{eq:relaxation_time}
\end{equation}
The remaining input to the equations of motion are the equation of state and the temperature dependence of the shear and bulk viscosities.

\begin{figure}
\includegraphics[width=7cm]{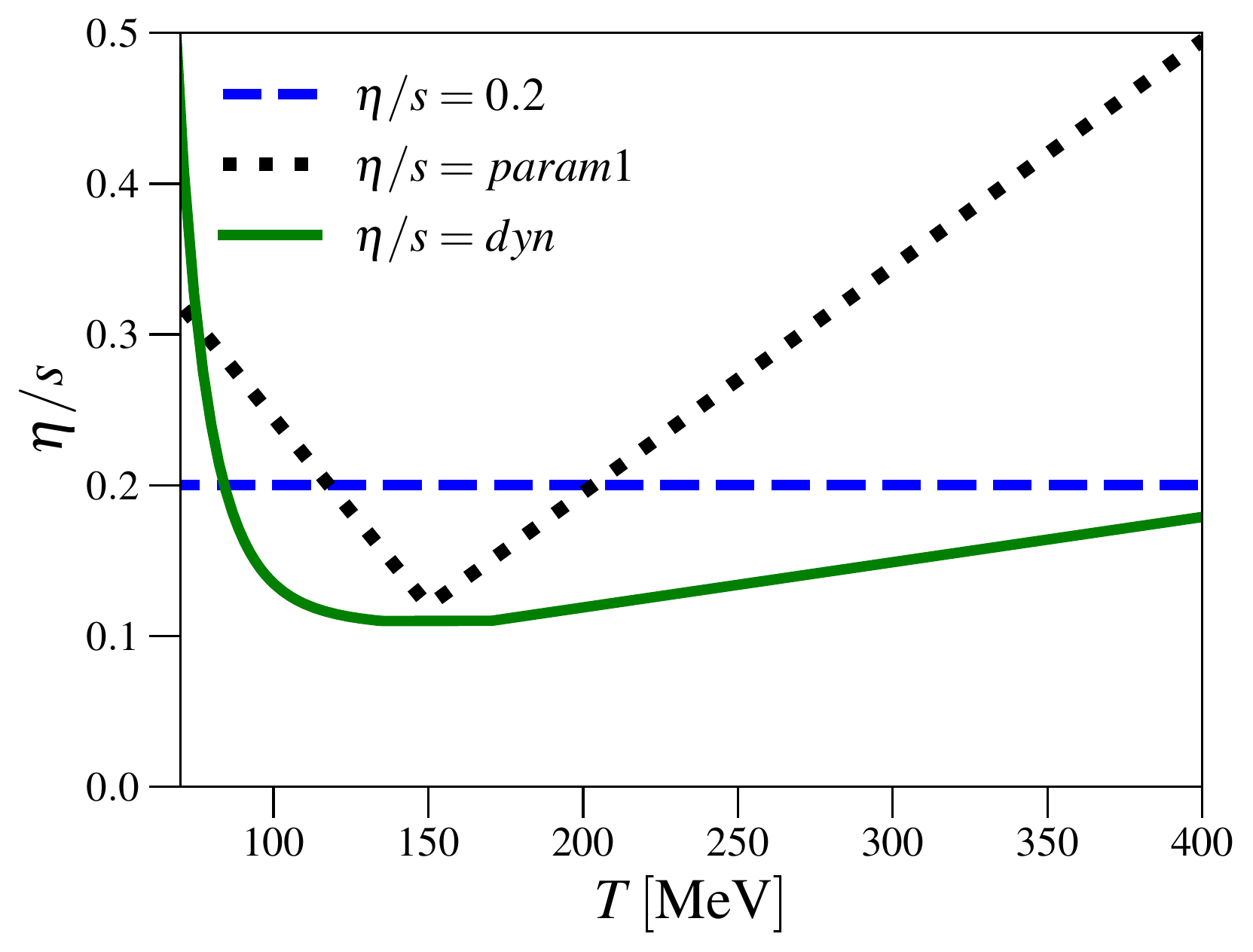}
\caption{(Color online) Shear viscosity to entropy density ratio as a function of temperature.}
\label{fig:etapers}
\end{figure}

\begin{figure}
\includegraphics[width=7cm]{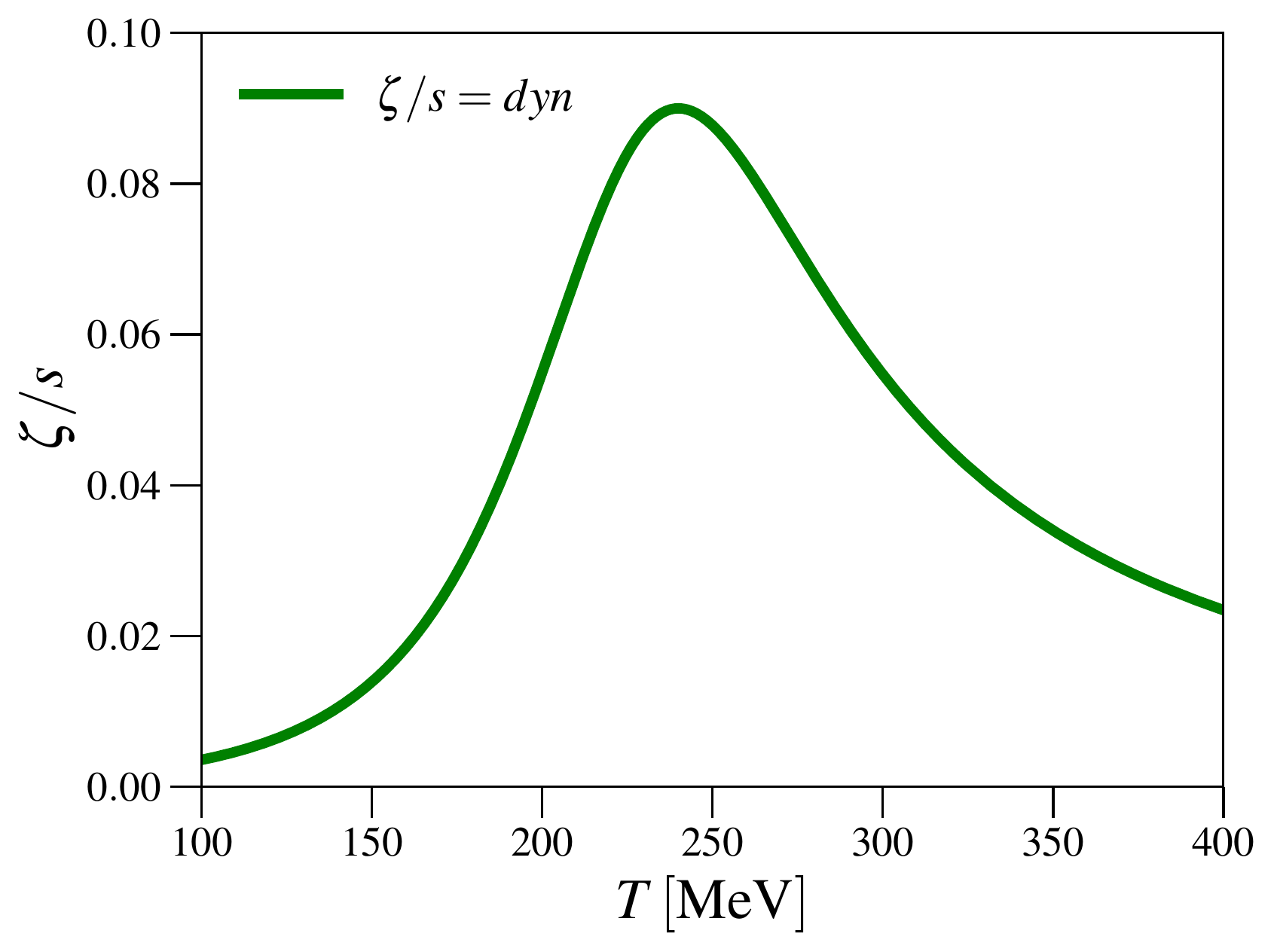}
\caption{(Color online) Bulk viscosity to entropy density ratio as a function of temperature.}
\label{fig:zetapers}
\end{figure}

The parametrizations of the shear viscosity to entropy density ratio are shown in Fig.~\ref{fig:etapers}, where $\eta/s=0.20$ and $\eta/s = param1$ are the same as implemented in earlier works \cite{Niemi:2015qia, Niemi:2015voa, Eskola:2017bup}. The new parametrization $\eta/s = dyn$ has a similar linear QGP part as the previous parametrizations while the hadronic part follows a power law, with power $P_H$, reaching its minimum $(\eta/s)_{min}$ at temperature $T_H$ followed by a constant part with width $W_{min}$, i.e. 

\begin{equation}
\eta/s (T) = 
\begin{cases}
(\eta/s)_{min} + S_{H}T\left(\Big(\frac{T}{T_H} \Big)^{-P_{H}} -1\right),  T<T_H\\
(\eta/s)_{min}, \hfill T_H\leq T \leq T_Q\\
(\eta/s)_{min} + S_{Q}(T-T_Q), \hfill  T>T_Q,
\end{cases}
\end{equation}
where $S_{H}$ and $S_{Q}$ are the slope parameters below $T_H$ and above $T_Q=T_H+W_{min}$, respectively.  The bulk viscosity is included together with the new $\eta/s = dyn$ parametrization and its ratio to entropy density is plotted as a function of temperature in Fig.~\ref{fig:zetapers}. Formally our paramaterization is written in a form:

\begin{equation}
    \zeta/s (T) = \frac{(\zeta/s)_{max}}{1+\big(\frac{T-T^{\zeta/s}_{max}}{w(T)}\big)^2},
\end{equation}
\begin{equation}
    w(T) = \frac{2 (\zeta/s)_{width}}{1+\exp\big({\frac{a_{\zeta/s}(T-T^{\zeta/s}_{max})}{(\zeta/s)_{width}}}\big)},
\end{equation}
where $(\zeta/s)_{max}$, $T^{\zeta/s}_{max}$, $(\zeta/s)_{width}$ and $a_{\zeta/s}$ are free parameters. The asymmetry parameter $a_{\zeta/s}$ describes the asymmetry of the bulk viscosity peak in such a way that $a_{\zeta/s}=0$ gives a completely symmetric peak. For the EoS we use the s95p parametrization~\cite{Huovinen:2009yb} of the lattice QCD results that includes the chemical freeze-out, implemented as effective chemical potentials in the hadronic part of the EoS~\cite{Bebie:1991ij, Hirano:2002ds, Huovinen:2007xh}. The earlier $\eta/s=0.20$ and $\eta/s = param1$ parametrizations use chemical freeze-out temperature $T_{\rm chem}=175$ MeV while the $\eta/s = dyn$ parametrization uses $T_{\rm chem}=155$ MeV. 

The transverse momentum spectra of hadrons are obtained by computing the Cooper-Frye freeze-out integrals on the kinetic decoupling surface for the hadrons included in the hadronic part of the EoS. The 2- and 3-body decays of unstable hadrons are accounted for. For the earlier parametrizations $\eta/s=0.20$ and $\eta/s = param1$ the kinetic decoupling surface is set to a constant $T_{\rm dec}=100$ MeV temperature hypersurface while the  $\eta/s = dyn$ parametrization uses dynamical criteria (see Sec. \ref{sec:freeze-out} for details) to determine the decoupling surface. The Cornelius algorithm~\cite{Huovinen:2012is} is employed to find the decoupling surface. The viscous correction $\delta f_i$ to each single-particle equilibrium momentum distribution, needed in the Cooper-Frye integrals, is implemented as in Refs.~\cite{Hosoya:1983xm, Gavin:1985ph, Sasaki:2008fg, Bozek:2009dw},

\begin{equation}
\begin{split}
\delta f_i =& -f_{0i}\tilde{f}_{0i} \frac{C_{bulk}}{T}\bigg[\frac{m^2}{3 E_k}-\Big(\frac{1}{3}-c_s^2\Big) E_k \bigg] \Pi \\
&+\frac{f_{0i}\tilde{f}_{0i}}{2T^2(e + P_0)} \pi^{\mu\nu} k_{\mu} k_{\nu},
\end{split}
\end{equation} 
where $k^{\mu}$ is the four-momentum of a given hadron, $E_k = u^\mu k_\mu$ is the energy of the hadron in the local rest frame, $f_{0i}$ is its equilibrium distribution, and $\tilde{f}_{0i} = 1 \pm f_{0i}$, with $+(-)$ for bosons (fermions). The coefficient $C_{bulk}$ is determined from
\begin{equation}
\frac{1}{C_{bulk}}= \sum_i \frac{g_i m_i^2}{3T} \int \frac{\mathrm{d}^3\mathbf{k}}{(2\pi)^3 k^0} f_{0i}\tilde{f}_{0i}\bigg[\frac{m_i^2}{3E_k}-\Big(\frac{1}{3}-c_s^2\Big)E_k\bigg].
\end{equation} 
Here $g_i$ is the degeneracy factor of a given hadron species $i$, and the sum includes all the species in the EoS. 

The fluid dynamical evolution and the transverse momentum spectra are computed for each collision event. The events are then grouped to the centrality classes according to the final charged particle multiplicities. However, if the experiments report the centrality of the collision by using the number of wounded nucleons, we can compute it by using the geometric collision criterion detailed at the end of Sec.~\ref{sec:ini}.

\begin{table}[]
\begin{center}
\begin{tabular}{||p{2.5cm} | p{1cm}||} 
\hline
\multicolumn{2}{||c||}{Initial state:}\\
\hline
$K_{\rm sat}$ & 0.67 \\
\hline
\hline
\multicolumn{2}{||c||}{Shear viscosity:}\\
\hline
$(\eta/s)_{min}$ & 0.11 \\
\hline
$T_H$ [ MeV ] & 135 \\
\hline
$S_{H}$ [ $\mathrm{GeV^{-1}}$ ] & 0.025 \\
\hline
$S_{Q}$ [ $\mathrm{GeV^{-1}}$ ] & 0.3 \\
\hline
$W_{min}$ [ MeV ] & 35\\
\hline
$P_{HG}$ & 8.0 \\
\hline
\multicolumn{2}{||c||}{Bulk viscosity:}\\
\hline
$(\zeta/s)_{max}$ & 0.09\\
\hline
$(\zeta/s)_{width}$ [ MeV ] & 60\\
\hline
$T_{max}^{\zeta/s}$ [ MeV ] & 240\\
\hline
$a_{\zeta/s}$ & -0.5 \\
\hline
\hline
\multicolumn{2}{||c||}{Dynamical freeze-out:}\\
\hline
$C_{Kn}$ & 0.8\\
\hline
$C_{R}$ & 0.15\\
\hline
\end{tabular}
\end{center}
\caption{Numerical values of the fit parameters used in the current study.}
\label{tab:etapers}
\end{table}

Numerical values of the parameters used here for the $\eta/s = dyn$ parametrization are shown in Table \ref{tab:etapers}. The initial state parameter $K_{\rm sat}$ is tuned to produce the same charged particle multiplicity in $2.76$ TeV Pb+Pb collisions as obtained in the ALICE measurements. Parameters of the shear viscosity and the dynamical freeze-out are iteratively adjusted to obtain results that match with ALICE measurements of $v_n\{2\}$ in $2.76$ TeV Pb+Pb collisions. Further tuning of the hadronic part of the $\eta/s$ parametrization is done to also match STAR measurements of $v_n\{2\}$ in central to mid-central $200$ GeV Au+Au collisions. The chemical freeze-out temperature is adjusted together with the parameters of bulk viscosity to achieve a good simultaneous agreement of the pion average $p_T$ and the proton multiplicity. 

We note that the idea here is that bulk viscosity in hadronic evolution is mainly described by chemical freeze-out~\cite{Paech:2006st,Dusling:2011fd,Rose:2020lfc}. In chemical freeze-out the corresponding bulk relaxation time is formally infinite, or at least much longer than the evolution time of the system, and the dynamics of the bulk pressure related to the non-equilibrium chemistry in this case cannot be readily computed using Israel-Stewart type of theory that assumes that the relaxation times are smaller than the evolution timescale. Instead, the bulk viscosity that is parametrized here should be thought as the residual bulk viscosity that is not included in the partial chemical freeze-out formalism~\cite{Huovinen:2007xh}. In practice, the condition that low-temperature bulk viscosity is described mainly by chemical freeze-out is set by adjusting the asymmetry parameter $a_{\zeta/s}$ in the parametrization such that bulk viscosity over entropy density becomes very small near and below the chemical freeze-out temperature. 

We want to emphasize here that this is only one example parametrization which seems to give a good agreement with the LHC and RHIC measurements. To get more detailed estimates of the parameters and their errors and correlations, a global analysis of heavy-ion observables and the parameter space is needed.

\section{Dynamical freeze-out}
\label{sec:freeze-out}
When modeling heavy-ion collisions using hydrodynamics the kinetic freeze-out is usually set to take place at a constant-temperature hypersurface. The basic argument is that the fluid decouples into free particles when the temperature dependent mean free path of the particles becomes of the same order as the size of the system $R$, i.e.\  $\lambda_{\rm mfp}(T) \sim R$. If the system size was a constant, this condition would give a constant freeze-out temperature. However, in reality the system size changes as a function of time, and moreover it can differ significantly from collision to collision. In particular, the systems created in central collisions are much larger than the ones created in peripheral collisions. 

A typical way to solve this issue is to connect fluid dynamics to a microscopic hadronic afterburner that automatically takes care of the freeze-out. However, a drawback in this approach is that it can easily lead to unphysical discontinuities in the transport coefficients, as 
at typical temperatures at the switching between fluid dynamics and hadron cascade the $\eta/s$ values in the fluid evolution are $\mathcal{O}(0.1)$, whereas on the hadron cascade side they are $\mathcal{O}(1)$~\cite{Rose:2017bjz, Prakash:1993bt, Csernai:2006zz}. Instead of a coupling to hadron cascade, in this work we treat the whole evolution, including the hadronic phase, using fluid dynamics. This has the specific advantage that it allows us to keep all the transport coefficients continuous throughout the whole temperature range realized in the evolution. 

In order to account for the nontrivial system size dependence of the freeze-out, we determine the decoupling surface dynamically \cite{Eskola:2007zc, Ahmad:2016ods} using two different conditions. The applicability of fluid dynamics requires that the local Knudsen number is sufficiently small, and fluid evolution becomes effectively free streaming when ${\rm Kn} \gg 1$. In comparisons between kinetic theory and fluid dynamics it was shown that a constant Knudsen number freeze-out in fluid dynamics catches very well the freeze-out dynamics of the kinetic evolution~\cite{Gallmeister:2018mcn}. On the other hand, even if the local condition gives that fluid dynamics is applicable, the overall size of the system can still be small compared to the mean free path of the particles. In order to account for this kind of nonlocal freeze-out, we impose a second condition that the fluid element decouples when the mean free path is of the same order as the system size. Hence, our dynamical freeze-out setup is determined by the following two conditions:

\begin{eqnarray}
 \mathrm{Kn} &=& \tau_{\pi} \theta = C_{\mathrm{Kn}} \\
 \label{eq:freeze1}
 \frac{\gamma\tau_{\pi}}{R} &=& C_{R},
 \label{eq:freeze2}
\end{eqnarray}
where $C_{\mathrm{Kn}}$, $C_R \sim 1 $ are some proportionality constants and $R$ is the size of the system. Here we have assumed that the mean free path is proportional to the relaxation time. The additional gamma factor in the second equation takes into account that the size of the system is calculated in the center-of-momentum frame of the nuclear collision, while the relaxation time is calculated in the fluid rest frame. To make sure that we are not in the QGP phase when freeze-out happens we also require that at the freeze-out surface $T<150$ MeV. In order to use latter condition (\ref{eq:freeze2}) we need to have some kind of estimate for the system size which, however, is not uniquely determined. In this work we define the size of the system as
\begin{equation}
    R=\sqrt{\frac{A}{\pi}},
\end{equation}
where $A$ is the area in the $x, y$ plane in which $\mathrm{Kn} < C_{\mathrm{Kn}}$. Additionally we take into account the possibility that the system may consist of multiple separate areas of a fluid and calculate the system size for each of these regions separately. We note that our approximation of the system size is close to the maximum length that a particle must travel from the center to the edge of the system. In practice, however, most of the matter is distributed closer to the edges of the system and most of the particles are moving with the fluid also towards the edge. For this reason the actual size of the system that the particles see can be significantly smaller than $R$, and as a result the proportionality constant $C_R$ can also be significantly smaller than 1.  

\begin{figure}
\includegraphics[width=8.5cm]{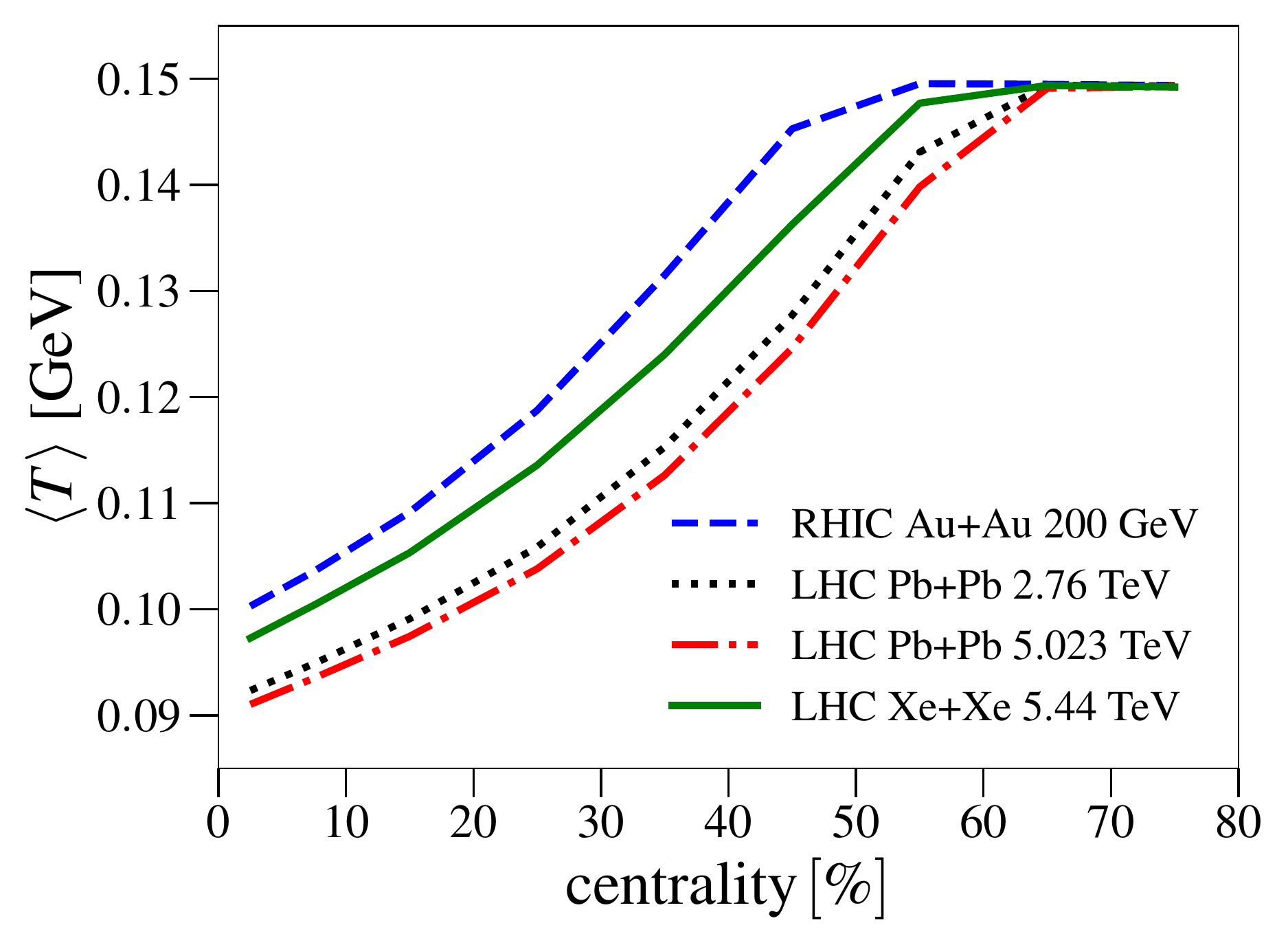}
\caption{(Color online) Average freeze-out temperature for $\eta/s = dyn$ parametrization in 200 GeV Au+Au, 2.76 TeV Pb+Pb, 5.023 TeV Pb+Pb, and 5.44 TeV Xe+Xe collisions.}
\label{fig:avetemp}
\end{figure}

In summary, here we have on the one hand reduced a possibly complicated non-equilibrium dynamics of the hadronic evolution in the dynamical treatment of kinetic freeze-out, and on the other hand we treat the non-trivial chemistry in the hadronic evolution as a constant-temperature chemical freeze-out. While such an approach may not catch the full microscopic details of the freeze-out dynamics, the purpose is that it would still capture its essential features. A clear advantage is, as mentioned above, that it allows us to keep the transport coefficients of the matter continuous throughout the evolution, and at the same time it also allows us to get constraints for the hadronic part of the transport coefficients. As we can see, the physical picture of the evolution is somewhat different from the typical hybrid hydro+cascade models, where the low viscosity QGP evolution is immediately followed by high-viscosity hadronic evolution. In our picture the peripheral collisions decouple practically immediately after the hadronization, but in the central collisions there can still be quite long low-viscosity evolution in the hadronic phase. This is demonstrated in Fig.~\ref{fig:avetemp} where the entropy-flux-weighted average freeze-out temperature is plotted as a function of centrality for the $\eta/s = dyn$ parametrization introduced in Sec.~\ref{sec:fluidsetup}. We can also notice that the average freeze-out temperature is sensitive to the collision energy and size of the colliding nuclei.

\section{Flow coefficients and correlators}
\label{sec:correlators}
The fluid dynamical computation gives a single-particle transverse momentum spectrum of hadrons for each event, and its azimuthal modulation can be expressed by its $p_T$ dependent Fourier components $v_n(p_T)$ and the phases or 
event-plane angles $\Psi_n(p_T)$,
\begin{equation}
\begin{split}
 \frac{dN}{dydp_T^2d\phi} \\ =
 \frac{1}{2\pi}&\frac{dN}{dydp_T^2}\left(1 + \sum_{n=1}^{\infty} v_n (p_T) \cos [n\left(\phi - \Psi_n(p_T)\right)] \right).
 \label{eq:fourier}
\end{split}
\end{equation}
The flow coefficients can be expressed in a convenient way by a complex flow vector $V_n$ as
\begin{equation}
 V_n(p_T) = v_n(p_T) e^{in\Psi_n(p_T)} = \langle e^{in\phi}\rangle_\phi,
\end{equation}
where the angular brackets denote an average
\begin{equation}
 \langle\cdots\rangle_\phi = \left(\frac{dN}{dydp_T^2}\right)^{-1} \int_0^{2\pi} d\phi  \frac{dN}{dydp_T^2d\phi} \left(\cdots\right).
\end{equation}
Similarly, the $p_T$-integrated flow coefficients can be defined as 
\begin{equation}
 V_n = v_n e^{in\Psi_n} = \langle e^{in\phi}\rangle_{\phi, p_T},
\end{equation}
where the average is defined as  
\begin{equation}
\langle \cdots \rangle_{\phi,p_T} = \left( \frac{dN}{dy}\right)^{-1}\int\limits_0^{2\pi} d\phi\!\! \int\limits_{\quad p_T, min}^{\quad p_T, max}\!\!\!\! dp_T^2 w \frac{dN}{dydp_T^2 d\phi}\left(\cdots\right),
\end{equation}
and the $p_T$-integrated multiplicity $\frac{dN}{dy}$ is defined with the same $p_T$ integration limits $p_ {T, \rm min}$ and $p_{T, \rm max}$ as above. In addition it is possible to use a $p_T$ or an energy dependent weight $w$ in the $p_T$ integration.

In the following we will write down the expressions of various measurable $p_T$-integrated quantities, but suppress the rapidity, weight, and $p_T$ integration limits from the notation. The $p_T$ limits will be denoted explicitly when we show our results. Unless otherwise stated, the weight function $w = 1$.

In the fluid dynamical simulations of heavy-ion collisions we are working directly with continuous particle distributions. In the experiments this is not the case, but each event is measured as a finite number of particles. Therefore, the definitions above are not directly applicable, but the flow coefficients are rather defined through particle correlations. As an example of a two-particle correlation and its continuum limit we can write
\begin{widetext}
\begin{equation}
 \frac{1}{N_e(N_e -1)} \sum_{{\rm pairs}\\ \, i \neq j} e^{in \phi_1} e^{-in \phi_2} \longrightarrow \frac{1}{N_2}\int d\phi_1 d\phi_2 \frac{dN_2}{d\phi_1 d \phi_2} e^{in \phi_1} e^{-in \phi_2},
\end{equation}
\end{widetext}
where $N_e$ is the number of hadrons in the event, and $dN_2/d\phi_1 d\phi_2$ is a two-particle distribution function that can be written as a sum of the product of the single-particle distribution functions and a direct correlation
\begin{equation}
 \frac{dN_2}{d\phi_1 d\phi_2} = \frac{dN}{d\phi_1} \frac{dN}{d\phi_2} + \delta_2(\phi_1, \phi_2),
\end{equation}
where the direct part emerges, e.g.\ due to hadron decays. It is a genuine two-particle correlation that is absent if all the correlations between the hadrons are due to the underlying collective flow. If the direct component can be neglected, the two-particle correlation above can be written in the continuum limit as
\begin{equation}
\begin{split}
 \frac{1}{N_2} \int d\phi_1 d\phi_2 \frac{dN}{d\phi_1} &\frac{dN}{d\phi_2} e^{in \phi_1} e^{-in \phi_2} \\
 &= v_n e^{in\Psi_n} v_n e^{-in\Psi_n} = v_n^2.
 \end{split}
\end{equation}
In this limit the two-particle correlator can be written in terms of the flow coefficient. This particular correlator is referred to as the two-particle cumulant, and its average over events gives the two-particle cumulant $v_n\{2\}$,
\begin{equation}
 v_n\{2\} = \sqrt{\langle v_n^2\rangle_{\rm ev}},
\end{equation}
where $\langle \cdots \rangle_{\rm ev}$ denotes the average over the events. A similar reasoning leads to a multitude of flow observables. Here we write down only the continuum limit in the absence of direct or nonflow correlations. It should be noted, however, that although the experimental procedures try to suppress the nonflow parts by e.g.\ requiring a rapidity gap between each pair of hadrons, it is still possible that some of the observables are still plagued by the non-flow. With the current setup we cannot address the non-flow part theoretically, but will assume that the experimental techniques remove them completely.

In a naive picture one may think that the flow coefficients are generated independently as a fluid dynamical response to the corresponding eccentricities of the initial conditions, $v_n \propto \varepsilon_n$. In practice, however, this picture holds only for the elliptic flow coefficient $v_2$ and to a lesser degree for $v_3$~\cite{Gardim:2011xv, Niemi:2012aj}, and even then the relation between $v_2$ and $\varepsilon_2$ ceases to be linear when $\varepsilon_2$ becomes large in noncentral collisions~\cite{Niemi:2015qia}. In general, the flow coefficients are not independent of each other, but both the correlations between the eccentricities in the initial conditions and the nonlinear fluid dynamical evolution generate correlations between them. The degree of the correlation can be measured through various observables that correlate both the magnitudes of the flow, $v_n$, and the event-plane angles $\Psi_n$~\cite{Gardim:2011xv}.

A measurable way to quantify the degree of correlation between the flow coefficients is the so called symmetric cumulant~\cite{ALICE:2016kpq}, defined as
\begin{equation}
 SC(n, m) = \langle v_n^2 v_m^2\rangle_{\rm ev, N^4} - \langle v_n^2\rangle_{\rm ev, N^2}\langle v_m^2\rangle_{\rm ev, N^2},
\label{eq:sc}
\end{equation}
where it is important to notice that the event-average is performed with powers of multiplicity as a weight, as denoted in the above equation. An advantage of this definition is that at the particle correlation level the latter term in the definition removes the direct two-particle correlations from the first term, which in turn is a four-particle correlator at the particle level. Thus the direct two-particle nonflow does not affect the symmetric cumulant. The symmetric cumulant is not a correlator in a sense that it depends not only on the degree of correlation between $v_n$ and $v_m$, but also on their absolute magnitudes. On the other hand, the normalized symmetric cumulant, defined as
\begin{equation}
 NSC(n, m) = \frac{SC(n,m)}{\langle v_n^2\rangle_{\rm ev, N^2}\langle v_m^2\rangle_{\rm ev, N^2}}
\label{eq:nsc}
\end{equation}
is a measure of only the correlation. The downside of the normalized version is that the normalization can be affected by the direct two-particle nonflow contributions.

The symmetric cumulants measure only correlations involving two second-order flow coefficients. The more general mixed harmonic cumulants ($MHC$) were introduced in Ref.~\cite{Moravcova:2020wnf} to give observables that can quantify the correlations between between more than two flow coefficients with higher-order moments of $v_n$'s. Like symmetric cumulants, mixed harmonic cumulants are also constructed in such a way that lower order correlations are removed from multiparticle correlations and the definition of $MHC$ containing two second order flow coefficients is identical to the symmetric cumulants, i.e.\ $MHC(v_m^2, v_n^2)=SC(v_m^2, v_n^2)$. Mixed harmonic cumulants for six-particle correlations involving moments of $v_2$ and $v_3$ can be defined as
\begin{equation}
\begin{split}
 MHC(v_2^4, v_3^2) =& \langle v_2^4 v_3^2 \rangle_{6} -4 \langle v_2^2 v_3^2\rangle_{4}\langle v_2^2\rangle_{2} \\
 &-\langle v_2^4\rangle_{4}\langle v_3^2\rangle_{2}+4 \langle v_2^2\rangle_{2}^2\langle v_3^2\rangle_{2} ,\\
 MHC(v_2^2, v_3^4) =& \langle v_2^2 v_3^4 \rangle_{6} -4 \langle v_2^2 v_3^2\rangle_{4}\langle v_3^2\rangle_{2} \\
 &-\langle v_2^2\rangle_{2}\langle v_3^4\rangle_{4}+4 \langle v_2^2\rangle_{2}\langle v_3^2\rangle_{2}^2,
\end{split} 
\label{eq:mhc6}
\end{equation}
where $\langle \cdots \rangle_i = \langle \cdots \rangle_{\rm ev, N^i}$. Similarly one can define mixed harmonic cumulants for eight-particle correlations between $v_2$ and $v_3$ as
\begin{equation}
\begin{split}
 MHC(v_2^6, v_3^2) =& \langle v_2^6 v_3^2 \rangle_{8} -9 \langle v_2^4 v_3^2\rangle_{6}\langle v_2^2\rangle_{^2} \\
 &-\langle v_2^6\rangle_{6}\langle v_3^2\rangle_{2}-9 \langle v_2^4\rangle_{4}\langle v_2^2 v_3^2\rangle_{4} \\
 &-36 \langle v_2^2\rangle_{2}^3\langle v_3^2\rangle_{2} +18 \langle v_2^2\rangle_{2}\langle v_3^2\rangle_{2} \langle v_2^4\rangle_{4},\\
 &+36\langle v_2^2\rangle_{2}^2\langle v_2^2v_3^2\rangle_{4},\\
 MHC(v_2^2, v_3^6) =& \langle v_2^2 v_3^6 \rangle_{8} -9 \langle v_2^2 v_3^4\rangle_{6}\langle v_3^2\rangle_{2} \\
 &-\langle v_2^2\rangle_{2}\langle v_3^6\rangle_{6}-9 \langle v_3^4\rangle_{4}\langle v_2^2 v_3^2\rangle_{4} \\
 &-36 \langle v_2^2\rangle_{2}\langle v_3^2\rangle_{2}^3 +18 \langle v_2^2\rangle_{2}\langle v_3^2\rangle_{2} \langle v_3^4\rangle_{4}\\
 &+36\langle v_3^2\rangle_{2}^2\langle v_2^2v_3^2\rangle_{4},\\
 MHC(v_2^4, v_3^4) =& \langle v_2^4 v_3^4 \rangle_{8} -4 \langle v_2^4 v_3^2\rangle_{6}\langle v_3^2\rangle_{2} \\
 &-4\langle v_2^2 v_3^4\rangle_{6}\langle v_2^2\rangle_{2}- \langle v_2^4\rangle_{4}\langle v_3^4\rangle_{4} \\
 &-8 \langle v_2^2 v_3^2\rangle_{4}^2 -24 \langle v_2^2\rangle_{2}^2\langle v_3^2\rangle_{2}^2\\
 &+4\langle v_2^2\rangle_{2}^2\langle v_3^4\rangle_{4}+4\langle v_2^4\rangle_{4}\langle v_3^2\rangle_{2}^2\\
 &+32\langle v_2^2\rangle_{2}\langle v_3^2\rangle_{2}\langle v_2^2 v_3^2\rangle_{4},
\end{split} 
\label{eq:mhc8}
\end{equation}
and for six-particle correlations between $v_2$, $v_3$ and $v_4$ as
\begin{equation}
\begin{split}
 MHC(v_2^2, v_3^2, v_4^2) =& \langle v_2^2 v_3^2 v_4^2\rangle_{6} - \langle v_2^2 v_3^2\rangle_{4}\langle v_4^2\rangle_{2}\\
 &-\langle v_2^2 v_4^2\rangle_{4}\langle v_3^2\rangle_{2}-\langle v_3^2 v_4^2\rangle_{4}\langle v_2^2\rangle_{2}\\
 & + 2\langle v_2^2\rangle_{2}\langle v_3^2\rangle_{2}\langle v_4^2\rangle_{2}.
\end{split}
\label{eq:mhc234}
\end{equation}
Analogously to normalized symmetric cumulants one defines normalized mixed harmonic cumulants as
\begin{eqnarray}
 nMHC(v_n^k, v_m^l) =& \frac{MHC(v_n^k, v_m^l)}{\langle v_n^k\rangle_{k} \langle v_m^l\rangle_{l}},\\
 nMHC(v_n^k, v_m^l, v_p^q) =& \frac{MHC(v_n^k, v_m^l, v_p^q)}{\langle v_n^k\rangle_{k} \langle v_m^l\rangle_{l} \langle v_p^q\rangle_{q}}.
\end{eqnarray}

A complementary observable to the symmetric cumulants, usually referred to as the event-plane correlator, is defined as~\cite{Luzum:2012da} 
\begin{multline}
 \langle \cos(k_1\Psi_1 + \cdots + n k_n\Psi_n)\rangle_{{\rm SP}} = \\
\frac{\langle v_1^{|k_1|} \cdots v_n^{|k_n|} \cos(k_1\Psi_1 + \cdots + n k_n\Psi_n)\rangle_{ev}}{\sqrt{\langle v_1^{2|k_1|} \rangle_{ev} \cdots \langle v_n^{2|k_n|} \rangle_{ev}}},
 \label{eq:epcorrelation}
\end{multline}
where the $k_n$'s are integers with the property $\sum_n n k_n = 0$ so that the correlator is independent of the azimuthal orientation. Despite its name it actually measures a correlation between both the magnitudes of the flow and event-plane angle, and in this sense provides complementary information to the symmetric cumulants above.

These correlations as such provide information that is independent from the flow magnitudes themselves, and give further independent constraints to the initial conditions and transport coefficients. However, it is interesting that the event-plane correlations are closely related to the magnitude of nonlinear response to the initial conditions~\cite{Yan:2015jma}. The basic idea in quantifying the nonlinear response is that the complex flow vector $V_n$ is divided into a linear part $V_{n\rm L}$ that is assumed to correlate only with the corresponding initial state eccentricity $\varepsilon_n$, and into a non-linear part that is independent of $\varepsilon_n$~\cite{Gardim:2011xv}. If we consider the simplest possible nonlinear 
contributions, we can write
\begin{eqnarray}
 V_4 &=& V_{4\rm L} + \chi_{4, 22} (V_2)^2 \\
 V_5 &=& V_{5\rm L} + \chi_{5, 23} V_2 V_3 \\
 V_6 &=& V_{6\rm L} + \chi_{6, 222} V_2^3 + \chi_{6, 33} V_3^2,
\end{eqnarray}
where $\chi$'s are the nonlinear response coefficients. Note that the nonlinear parts include only the largest flow vectors $V_2$ and $V_3$ that can also, to a reasonable approximation as discussed above, assumed to have only the linear part $V_2 = V_{2\rm L}$ and $V_3 = V_{3\rm L}$. If we further assume that the linear and nonlinear parts are uncorrelated, we may express the response coefficients as
\begin{eqnarray}
 \chi_{4, 22} &=& \frac{\Re\langle V_4 (V_2^*)^2 \rangle_{ev}}{\langle |V_2|^4 \rangle_{ev}} \\
 \chi_{5, 23} &=& \frac{\Re\langle V_5 V_2^* V_3^* \rangle_{ev}}{\langle |V_2|^2 |V_3|^2  \rangle_{ev}} \\
 \chi_{6, 222} &=& \frac{\Re\langle V_6 (V_2^*)^3 \rangle_{ev}}{\langle |V_2|^6 \rangle_{ev}} \\
 \chi_{6, 33} &=& \frac{\Re\langle V_6 (V_3^*)^2 \rangle_{ev}}{\langle |V_3|^4 \rangle_{ev}},
\end{eqnarray}
and the linear parts of $V_4$ and $V_5$ can be written as
\begin{eqnarray}
 \sqrt{\langle|V_{4\rm L}| \rangle_{ev}^2} & = & \sqrt{ (v_4\{2\})^2 - \chi_{4, 22}^2 \langle |V_2|^4 \rangle_{ev}  } \\
 \sqrt{\langle|V_{5\rm L}| \rangle_{ev}^2} & = & \sqrt{ (v_5\{2\})^2 - \chi_{5, 23}^2 \langle |V_2|^2 |V_3|^2\rangle_{ev}}.
\end{eqnarray}
The connection between the event-plane correlators and the nonlinear response coefficients can be seen by observing, e.g.\ that
\begin{equation}
 \chi_{4, 22} = \langle \cos(4[\Psi_4 - \Psi_2)\rangle_{{\rm SP}}\sqrt{\frac{\langle v_4^2\rangle_{ev}}{\langle v_2^4\rangle_{ev}}},
\end{equation}
so that the two measures differ by a normalization factor that depends on the magnitude of the flow, but not on correlators. A similar
connection can also be made between the other $\chi$'s. A more complete list of relations can be found from Refs.~\cite{Yan:2015jma, Giacalone:2016afq}.

Even though the nonlinear response coefficients and the correlations between the flow harmonics give information about the initial state eccentricities and their conversion to momentum space anisotropies, they do not directly probe the size of the initial nuclear overlap region which is more sensitive to the average $p_T$ fluctuations. Thus, the correlation between the flow coefficients and the average $p_T$ is a good probe of the initial state structure \cite{Giacalone:2020byk}. This flow-transverse-momentum correlation is defined by a modified Pearson correlation coefficient \cite{Schenke:2020uqq}
\begin{equation}
    \rho(v_n^2, [p_T]) = \frac{\langle \hat{\delta} v_n^2 \hat{\delta} [p_T]\rangle_{\rm ev}}{\sqrt{\langle (\hat{\delta} v_n^2)^2\rangle_{\rm ev}\langle (\hat{\delta} [p_T])^2\rangle_{\rm ev}}},
\end{equation}
where the event-by-event variance at a fixed multiplicity for some observable $O$ is defined by
\begin{equation}
    \hat{\delta}O = \delta O- \frac{\langle\delta O \delta N \rangle_{\rm ev}}{\sigma_N} \delta N,
\end{equation}
\begin{equation}
    \delta O = O - \langle O \rangle_{\rm ev}, \hspace{3mm} \sigma_O^2 = \langle (\delta O)^2 \rangle_{\rm ev}.
\end{equation}

\section{results}
\label{sec:results}

In this section we present the results for hadron multiplicities, average $p_T$, flow coefficients and correlations calculated from the EKRT pQCD + hydrodynamics framework with the bulk viscosity and the dynamical freeze-out, and compare these against the results from our earlier works~\cite{Niemi:2015qia, Niemi:2015voa, Eskola:2017bup, Eskola:2017imo} with the constant-temperature freeze-out and without the bulk viscosity. The systems we show here are 200 GeV Au+Au, 2.76 TeV Pb+Pb, 5.023 TeV Pb+Pb, and 5.44 TeV Xe+Xe collisions. As explained in Sec.~\ref{sec:fluidsetup}, the initial conditions, the transport coefficients and the freeze-out parameters are fixed on the basis of 200 GeV Au+Au and 2.76 TeV data from RHIC and LHC. For both Pb+Pb collision systems we run 40000 event simulations to get better statistics for the symmetric cumulants while for other collision systems we did 20000 event simulations. The statistical errors for different quantities are estimated, as in Ref.~\cite{Giacalone:2016afq}, via jackknife resampling.

\subsection{Multiplicity, average $p_T$ and flow}
\begin{figure}
\includegraphics[width=8.5cm]{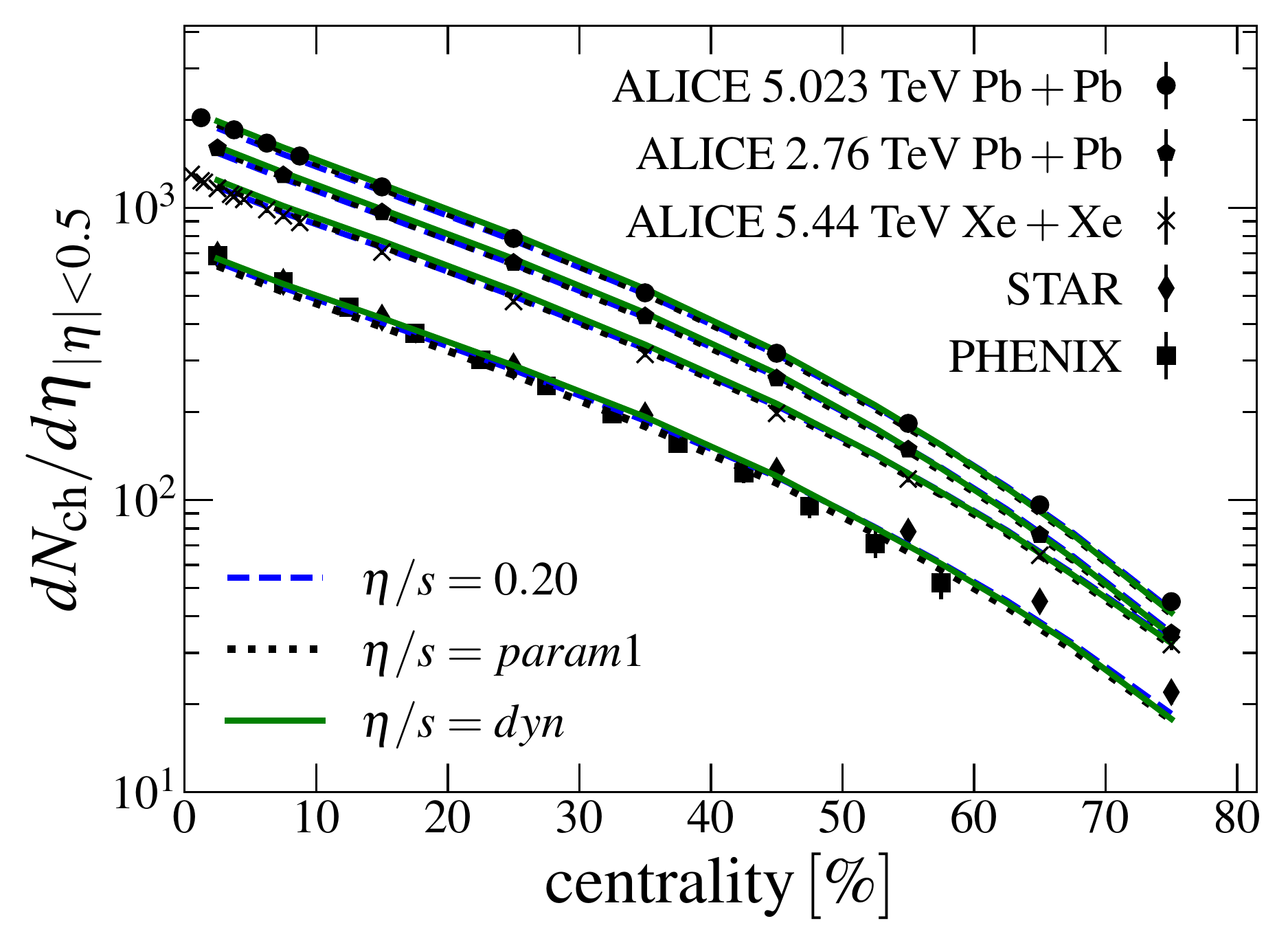}
\caption{(Color online) Charged hadron multiplicity in $200$ GeV Au+Au, $2.76$ TeV Pb+Pb, $5.023$ TeV Pb+Pb, and $5.44$ TeV Xe+Xe collisions.
The experimental data are from the ALICE \cite{Aamodt:2010cz, Adam:2015ptt, Acharya:2018hhy}, STAR \cite{Abelev:2008ab} and PHENIX \cite{Adler:2004zn} Collaborations.}
\label{fig:multiplicity}
\end{figure}

In Fig.~\ref{fig:multiplicity} we show the centrality dependence of charged hadron multiplicities for all the above systems compared to the STAR \cite{Abelev:2008ab}, PHENIX \cite{Adler:2004zn}, and ALICE \cite{Aamodt:2010cz, Adam:2015ptt, Acharya:2018hhy} data. The essential parameter that controls the multiplicity is $K_{\rm sat}$ in the local saturation criterion. This coefficient is fixed from the multiplicity in 0-5 \% 2.76 TeV Pb+Pb collisions. The centrality, $\sqrt{s_{\rm NN}}$, and nuclear mass number dependence are predictions of the model. The value of $K_{\rm sat}$ depends on the chosen $\eta/s(T)$ and $\zeta/s(T)$ parametrizations due to the different entropy production with different shear and bulk viscosities. However, the final results for the multiplicities are in practice the same for all parametrizations and they agree excellently with the experimental data across all centrality classes and collision energies. 
\begin{figure*}
\includegraphics[width=8.5cm]{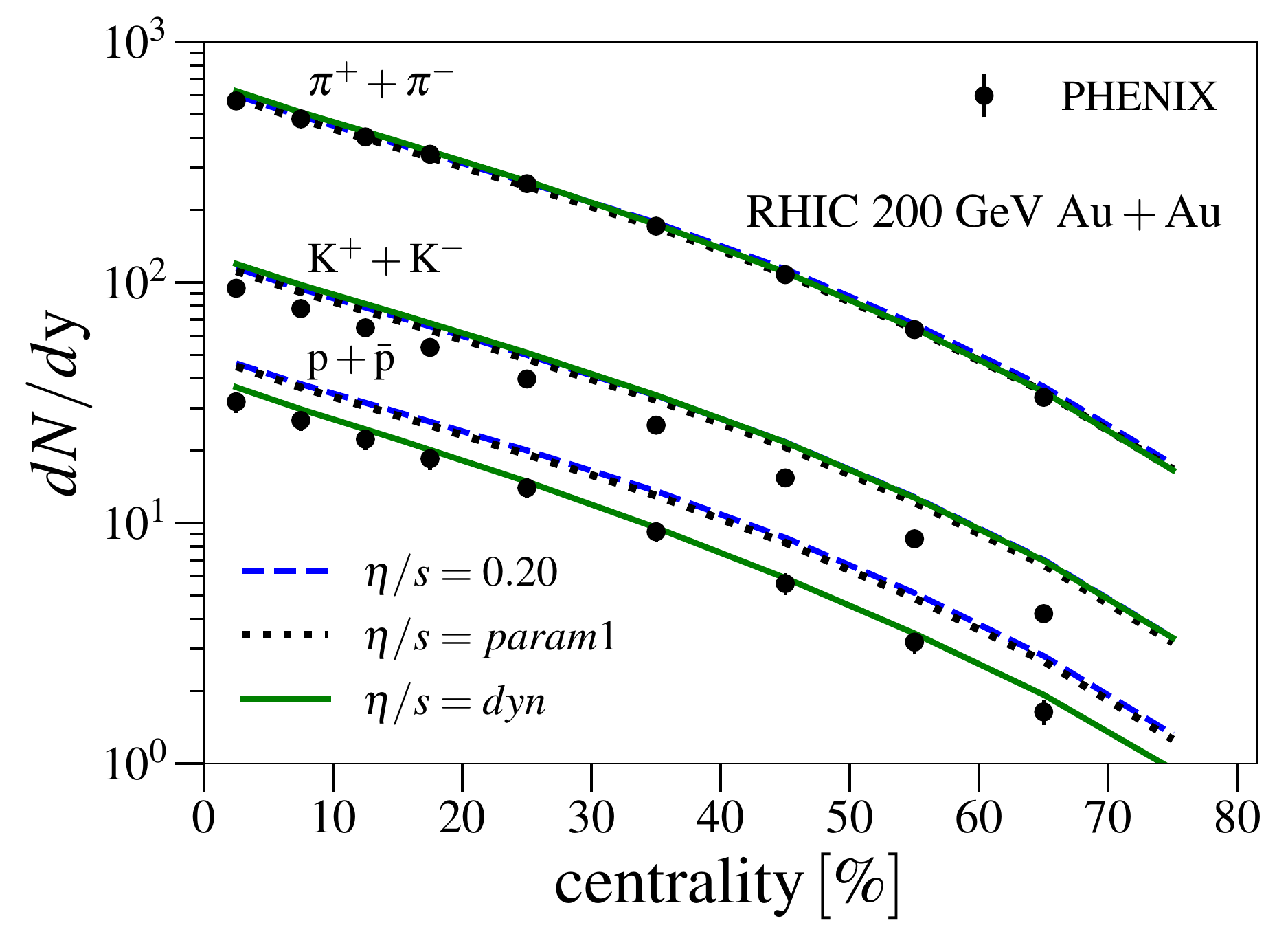}
\includegraphics[width=8.5cm]{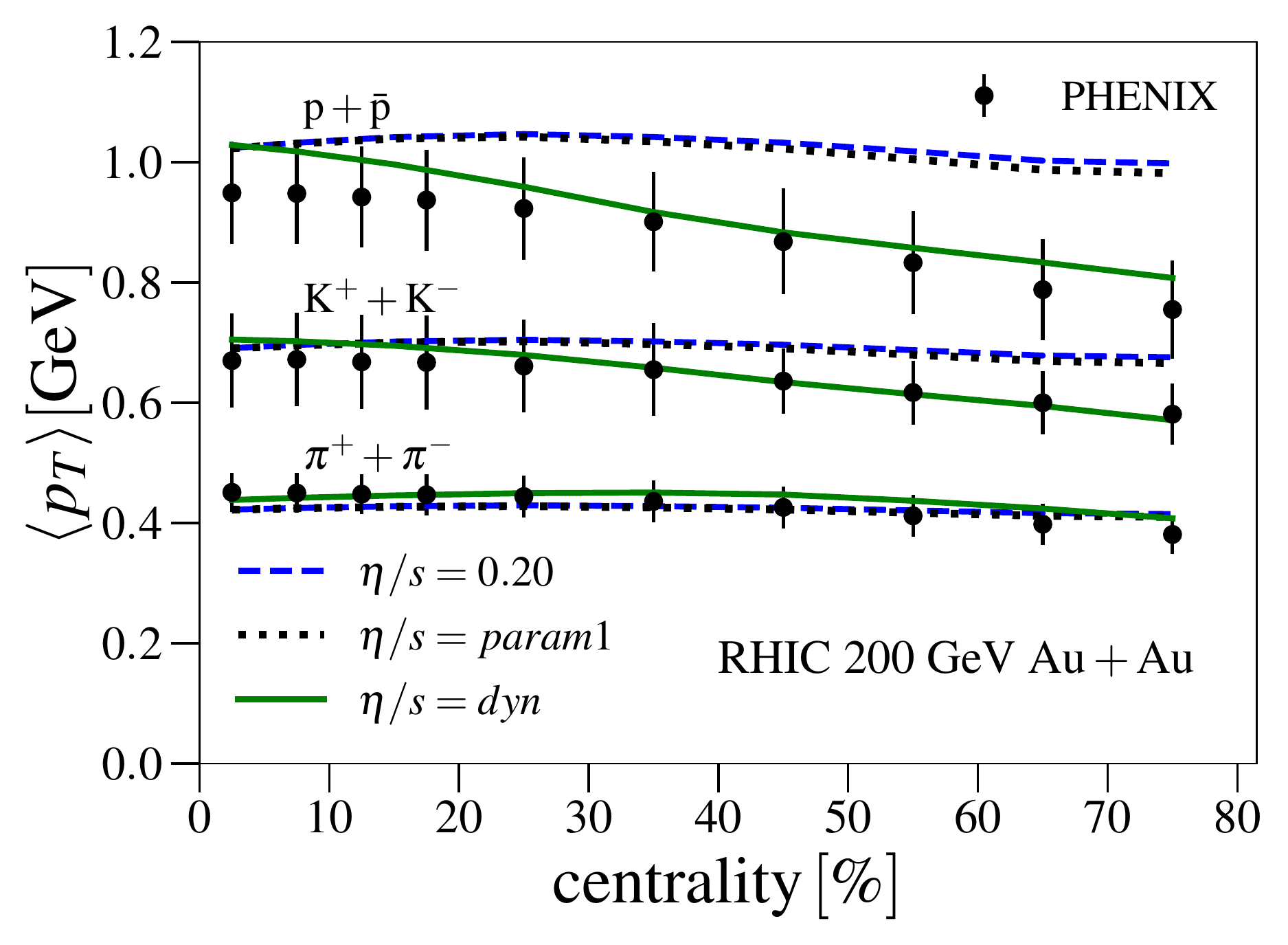}
\includegraphics[width=8.5cm]{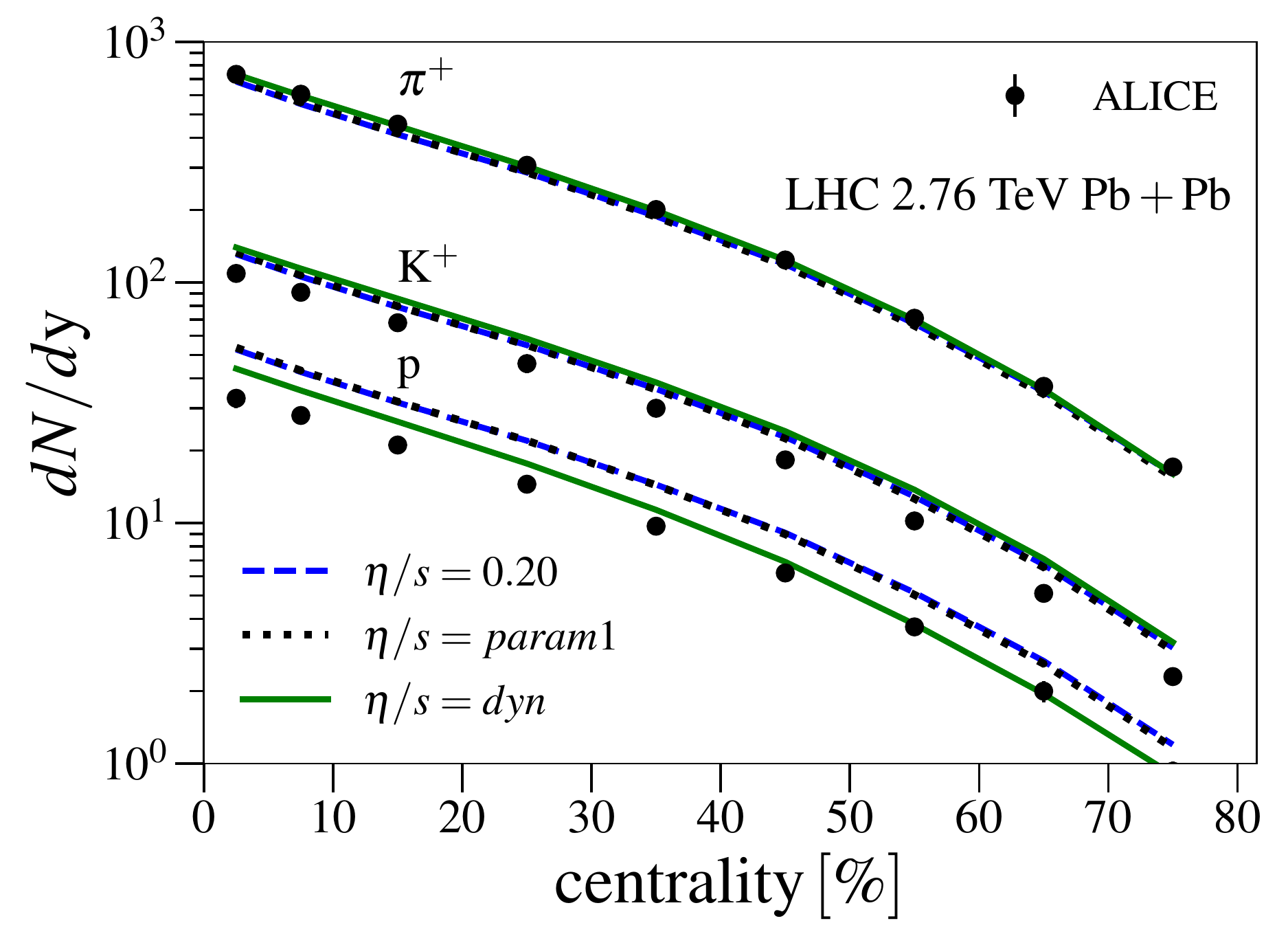}
\includegraphics[width=8.5cm]{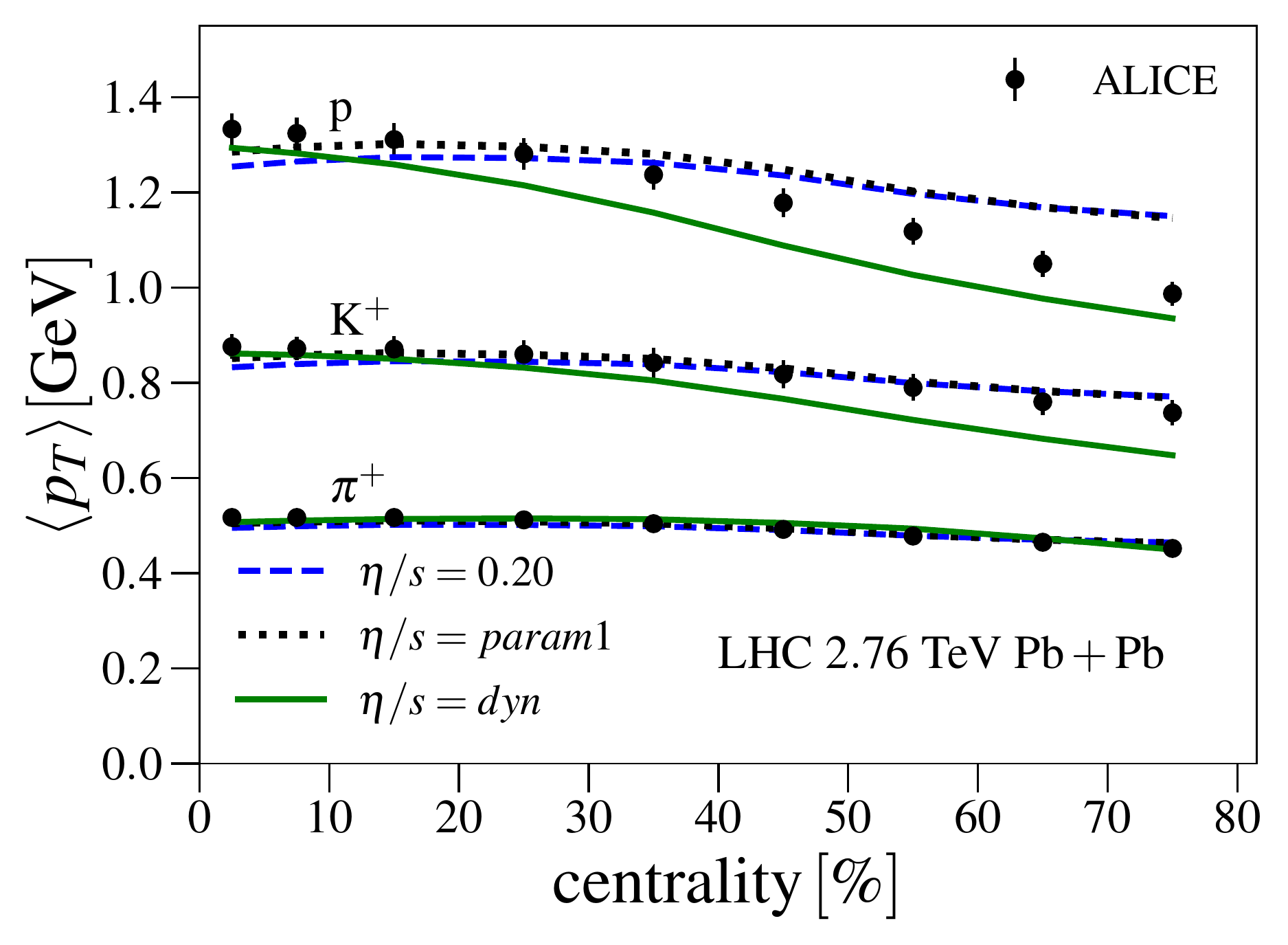}
\includegraphics[width=8.5cm]{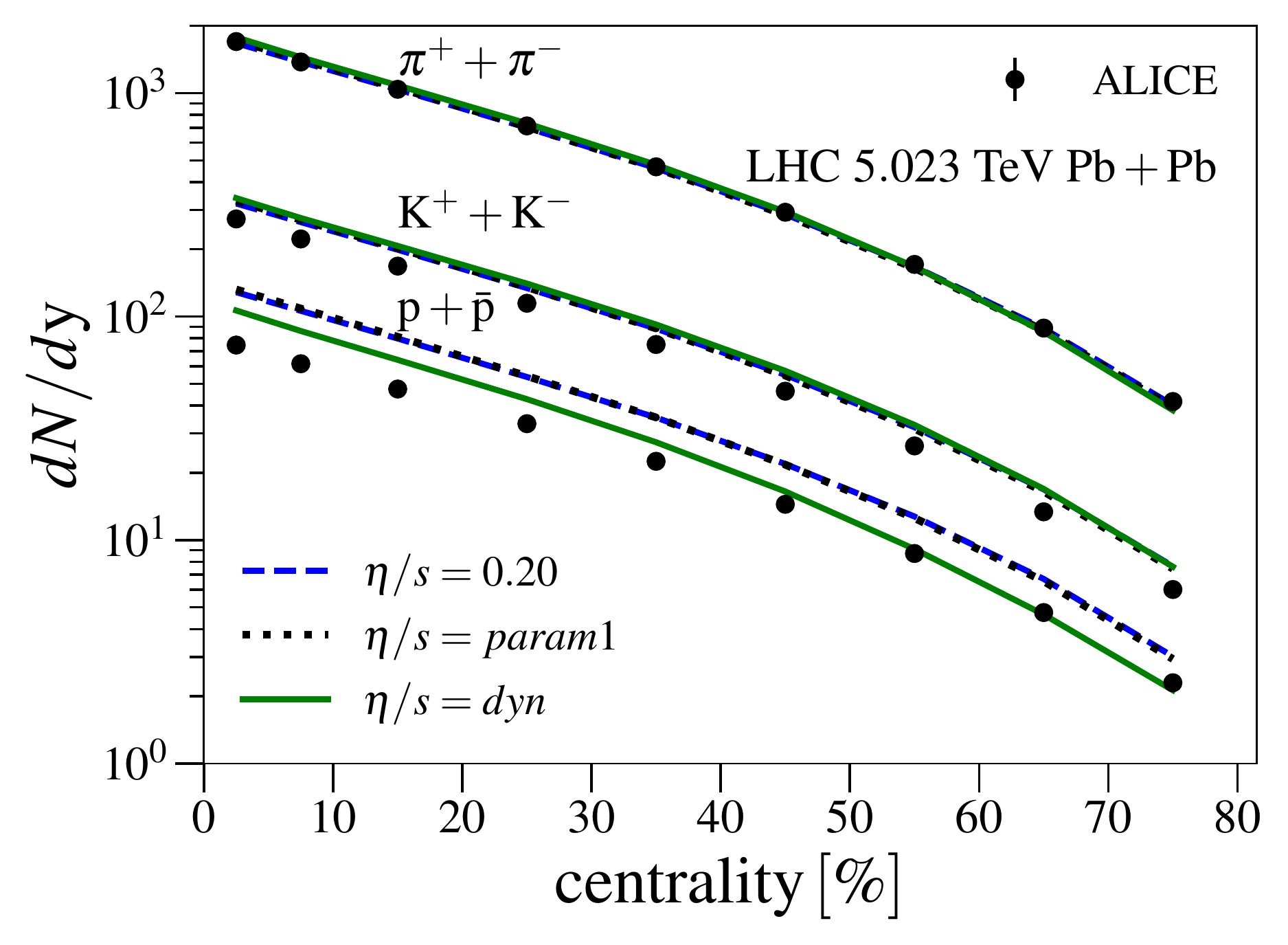}
\includegraphics[width=8.5cm]{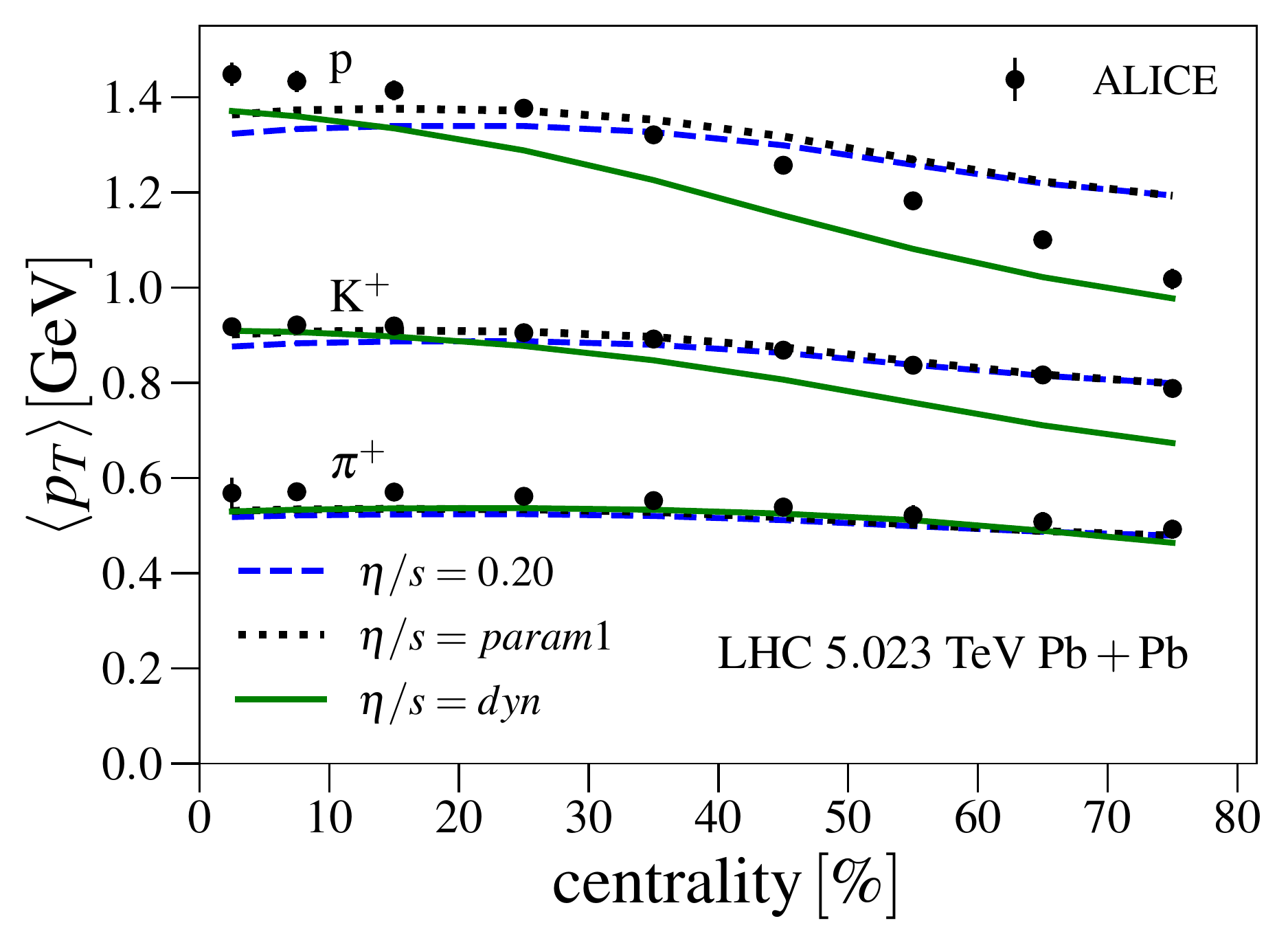}

\caption{(Color online) Identified particle multiplicities (left) and average transverse momenta (right) for pions, kaons and protons in $200$ GeV Au+Au, $2.76$ TeV Pb+Pb and $5.023$ TeV Pb+Pb collisions. The experimental data are from the PHENIX \cite{PHENIX:2003iij} and ALICE \cite{ALICE:2013mez,ALICE:2019hno} Collaborations.}
\label{fig:multiplicity_identified}
\end{figure*}

The centrality dependences of identified particle multiplicities for 200 GeV Au+Au 2.76 TeV Pb+Pb and 5.023 TeV Pb+Pb collisions are shown in Fig.~\ref{fig:multiplicity_identified} (left). All of the parametizations manage to produce the same pion multiplicities as the ALICE and PHENIX measurements while the kaon multiplicities differ significantly from the experimental data. The ratio between the proton and pion multiplicities is mostly controlled by the chemical freeze-out temperature. Parametrizations $\eta/s = 0.2$ and $\eta/s = param1$ use $T_{\rm chem} = 175$ MeV in order to obtain the same average $p_T$ for pions in 2.76 TeV Pb+Pb collisions as the ALICE measurements. However this comes with the drawback that the proton multiplicities differ from the experimental data by a factor of $\sim$ 2. The addition of the bulk viscosity in the $\eta/s=dyn$ parametrization enables the possibility to use $T_{\rm chem} = 155$ MeV which clearly improves the proton multiplicities. However, there is still some discrepancy left that is most visible in the most central collisions at the LHC.

In Fig.~\ref{fig:multiplicity_identified} (right) we show the average $p_T$ of identified particles as a function of centrality for 200 GeV Au+Au, 2.76 TeV Pb+Pb and 5.023 TeV Pb+Pb collisions. Compared to the earlier results, the $\eta/s=dyn$ parametrization improves the agreement with the experimental data across both collision systems, except for kaons at the LHC energies. In particular, the relative change of the proton $\langle p_T \rangle$  as a function of centrality is reproduced better. This improvement is not only due to the addition of the bulk viscosity but also the dynamical freeze-out plays a major part by affecting the lifetime of the fluid. 

\begin{figure*}
\includegraphics[width=8.5cm]{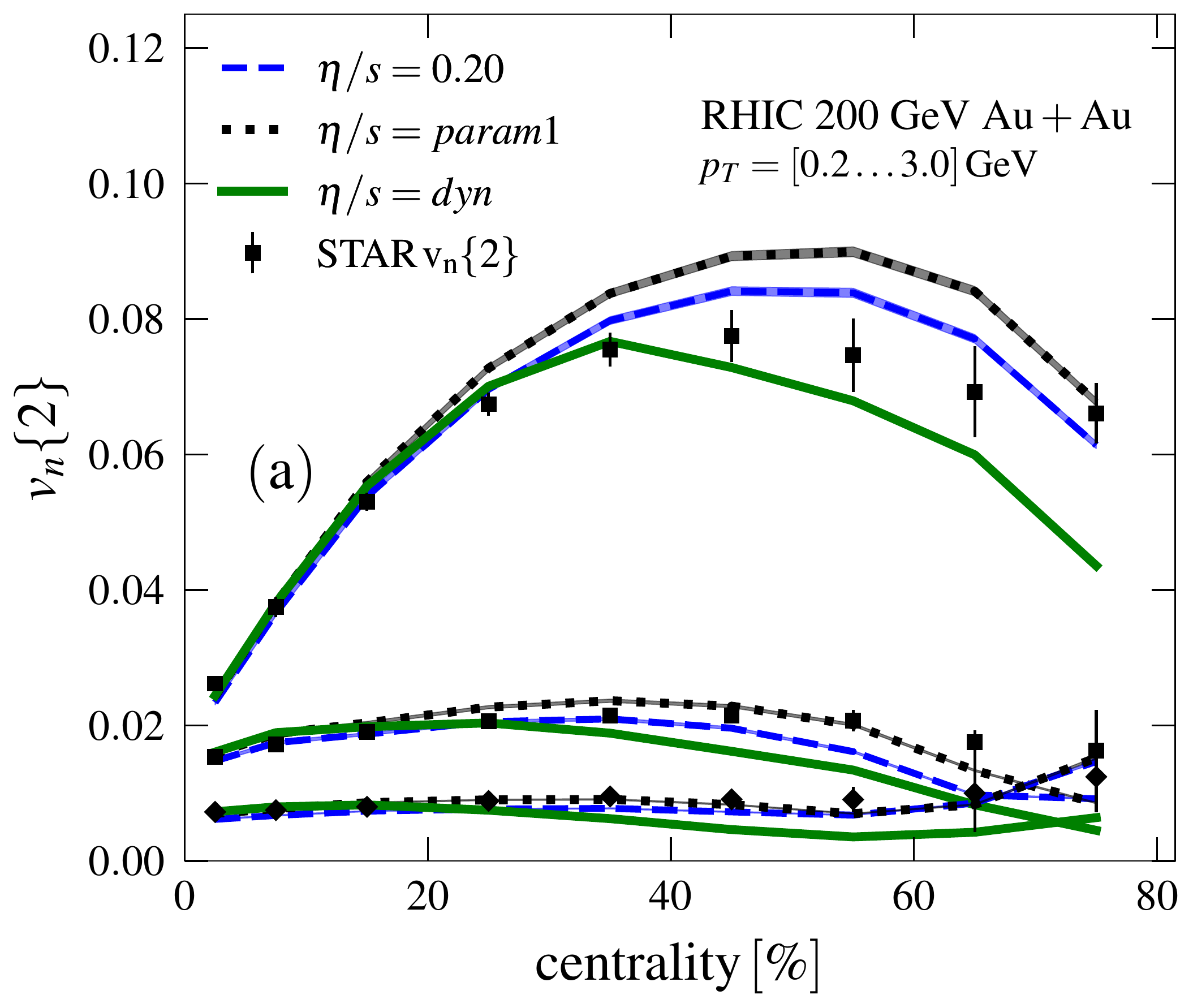}
\includegraphics[width=8.5cm]{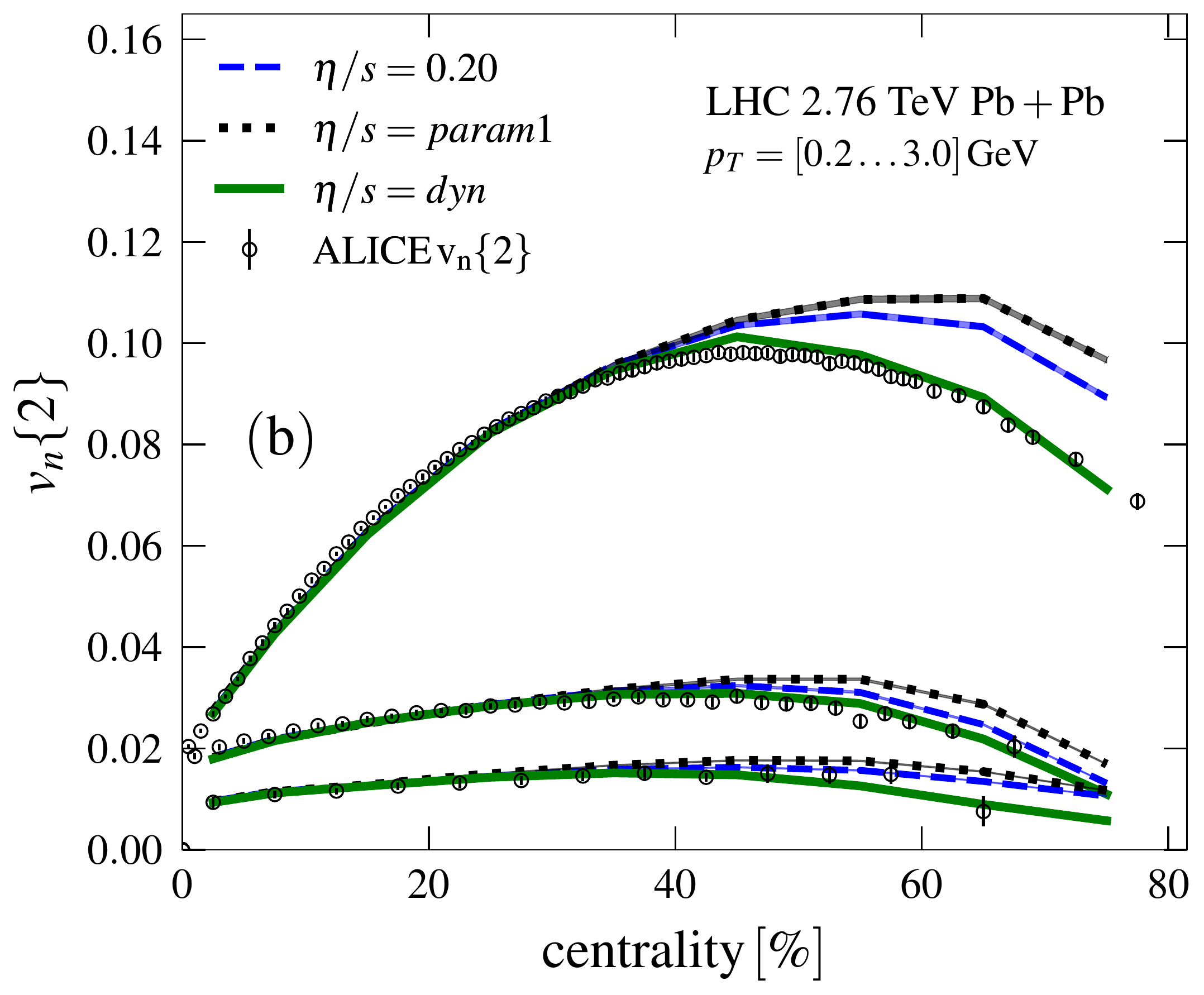}
\includegraphics[width=8.5cm]{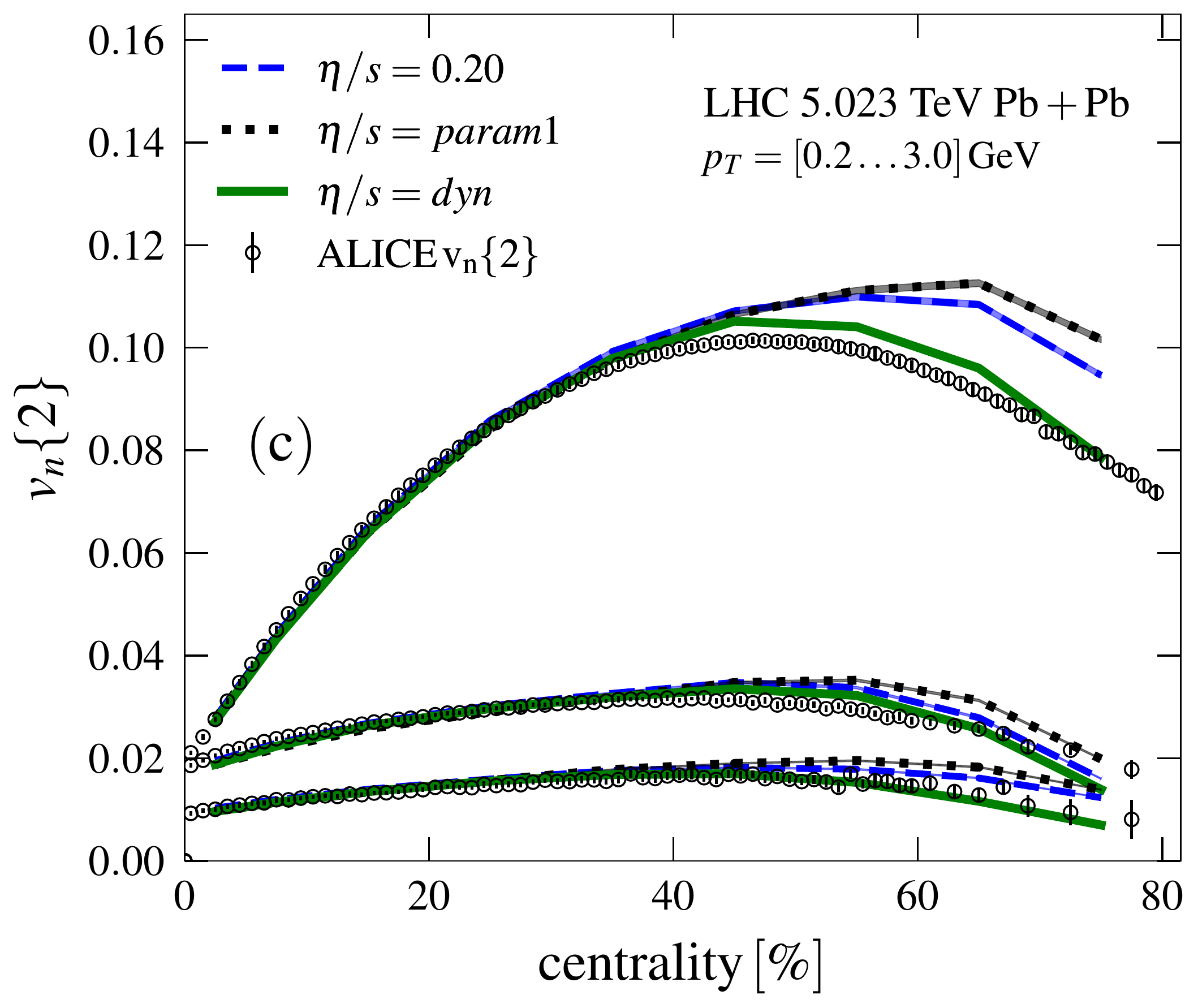}
\includegraphics[width=8.5cm]{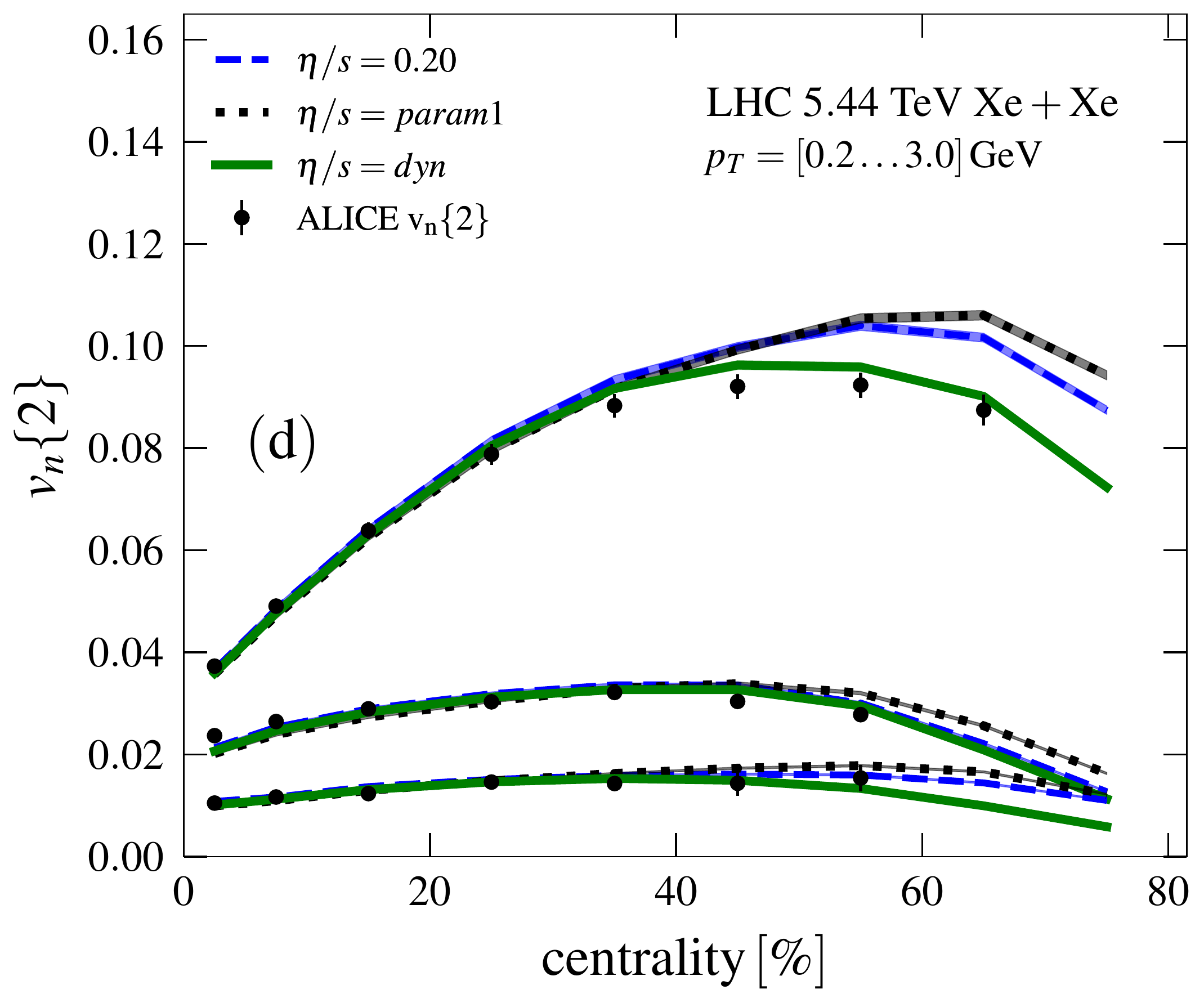}
\caption{(Color online) Flow coefficients in $200$ GeV Au+Au (a), $2.76$ TeV Pb+Pb (b), $5.023$ TeV Pb+Pb (c), and $5.44$ TeV Xe+Xe (d) collisions.
The experimental data are from the STAR \cite{STAR:2016vqt,STAR:2017idk} and ALICE Collaborations \cite{ALICE:2018rtz, ALICE:2018lao}.}
\label{fig:flow}
\end{figure*}
The centrality dependencies of the $p_T$-integrated flow coefficients $v_2\{2\}$, $v_3\{2\}$, and $v_4\{2\}$ in all studied systems are shown in Fig.~\ref{fig:flow}. The shear viscosity and the dynamical freeze-out parameters  of the $\eta/s = dyn$ parametrization were tuned to approximately reproduce $v_2\{2\}$ in $2.76$ TeV Pb+Pb collisions while also reproducing $v_2\{2\}$ in central to mid-central 200 GeV Au+Au collisions. The most essential feature of the dynamical freeze-out is that the smaller collision systems freeze out earlier in the hadronic phase. This means that there is less time for the initial state eccentricities to convert to the momentum space anisotropies in peripheral collisions. Indeed, as seen in Fig.~\ref{fig:flow}, all $p_T$-integrated flow coefficients for the $\eta/s = dyn$ parametrization are significantly smaller in peripheral collisions than the results of the $\eta/s$ parametrizations from the earlier works that used a constant-temperature decoupling surface. As can be seen from the comparison to measurements, the $\eta/s = dyn$ parametrization reproduces well the centrality dependence of all flow coefficients in all LHC collision systems and clearly improves the results from the earlier ones in peripheral collisions. The biggest discrepancy with the data and the model calculation is the $40-80\%$ -centrality range in 200 GeV Au+Au collisions. In this region especially the predictions for the flow coefficients $v_3\{2\}$ and $v_4\{2\}$ are well outside of the error bars of the measurements. There are multiple possible reasons for this. First of all, due to the lower multiplicity in the 200 GeV Au+Au collisions it is reasonable to expect significantly larger nonflow effects compared to the LHC systems. Additionally, the $\delta f$ corrections to the particle spectra are much larger at RHIC than at LHC which adds additional uncertainty to the RHIC results. Lastly, we do not include any nucleon substructure~\cite{Mantysaari:2016ykx}, initial flow or nonzero $\pi^{\mu\nu}$ to our initial state model and effects of these modifications are still under investigation. We note that other groups report very similar flow coefficients in peripheral RHIC collisions; see, e.g., Refs.~\cite{JETSCAPE:2020mzn, Gale:2021emg}

\begin{figure*}
\includegraphics[width=12.0cm]{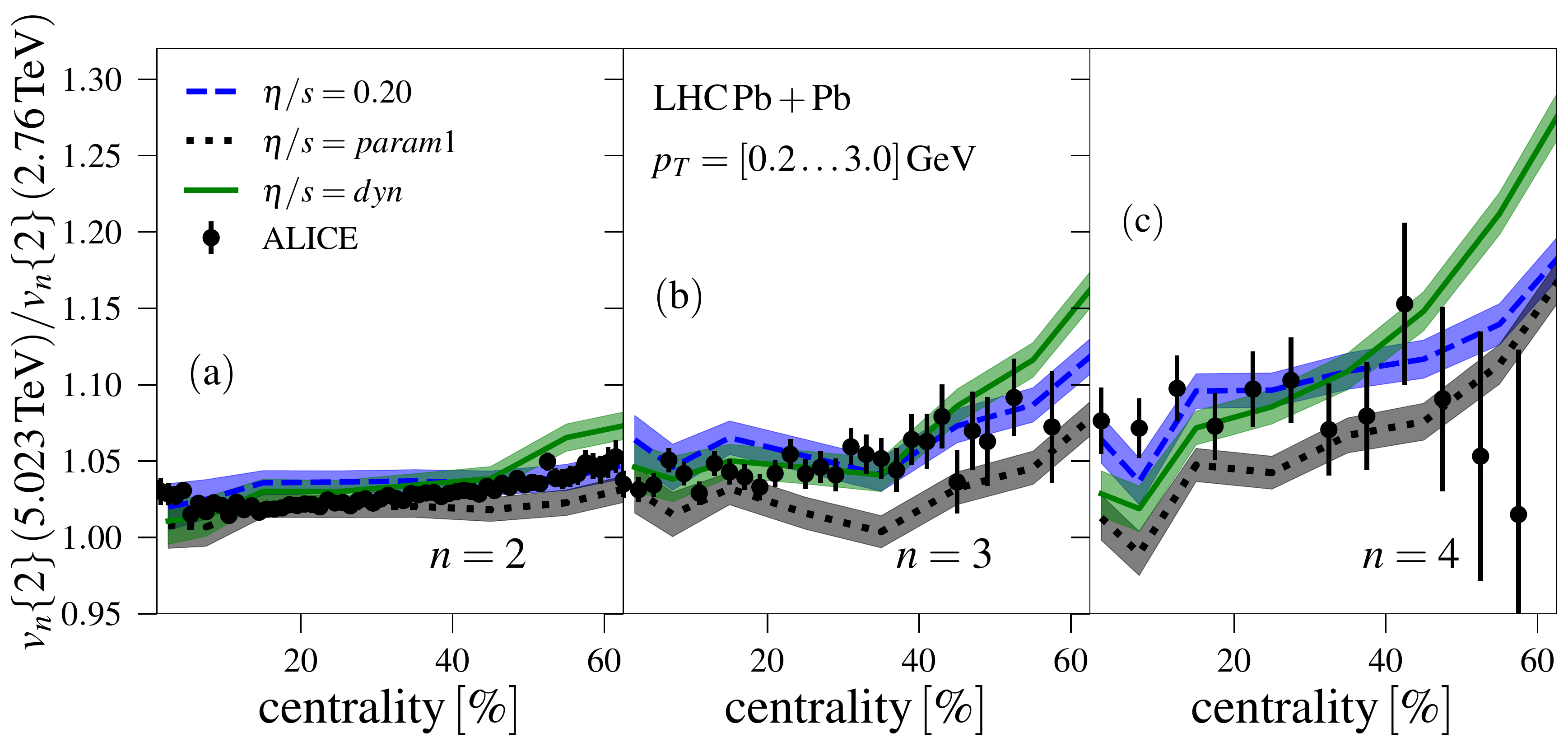}
\includegraphics[width=12.0cm]{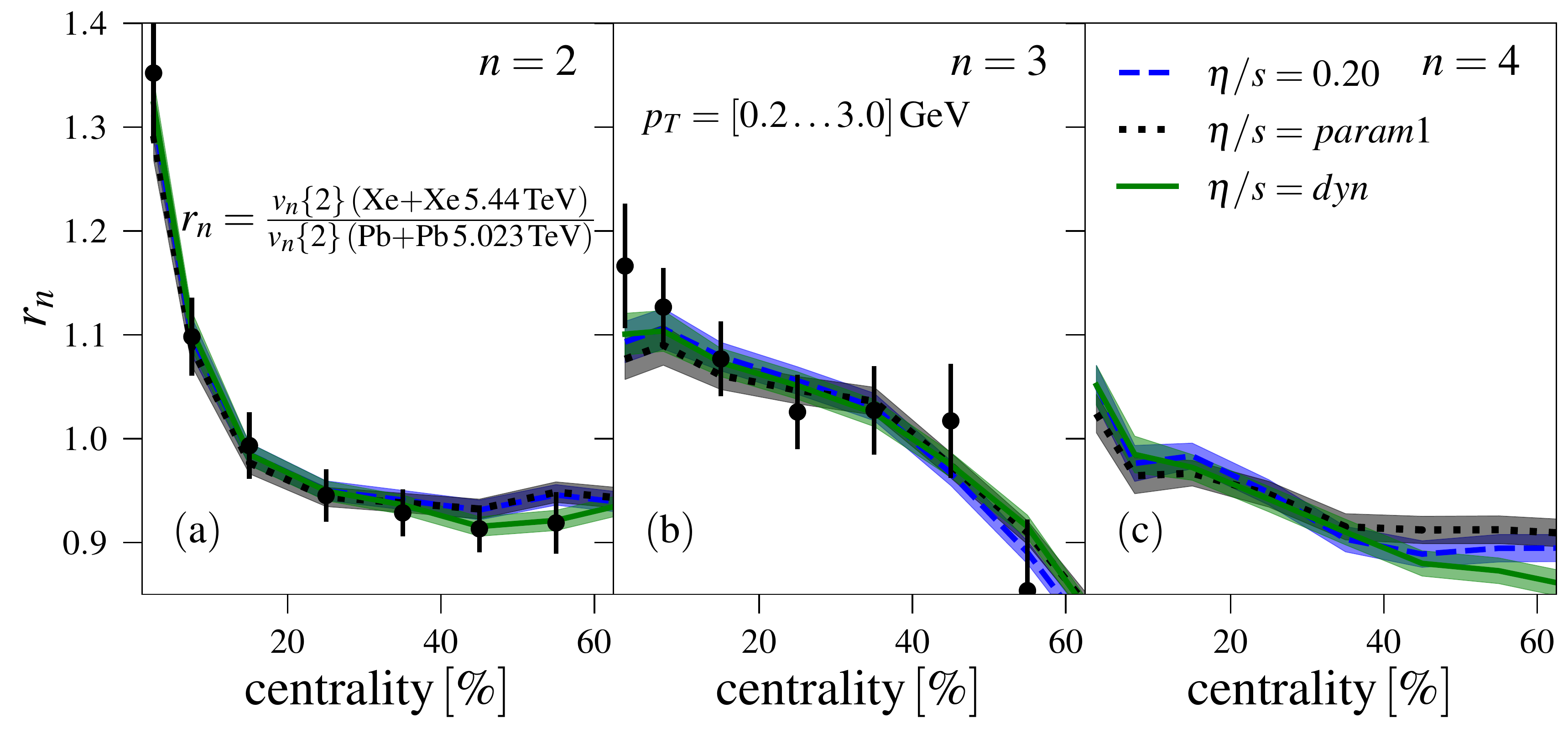}
\caption{(Color online) The ratio of flow coefficients $v_n$ between $5.023$ TeV and $2.76$ TeV Pb+Pb collisions, and the ratio of $v_n$ between
$5.44$ TeV Xe+Xe and $5.023$ TeV Pb+Pb collisions. The experimental data are from the ALICE Collaboration \cite{ALICE:2018rtz, ALICE:2018lao}.}
\label{fig:flowratio}
\end{figure*}

The change in the magnitude of the flow coefficients is quite modest from 2.76 to 5.023 TeV Pb+Pb collisions, and a better way to quantify the change is to plot the ratio of the coefficients between the two collision energies. The ratio is also a more robust prediction from fluid dynamics and less sensitive to
fine tuning of $\eta/s(T)$, for a discussion see Ref.~\cite{Noronha-Hostler:2015uye}. The predictions for the ratios of $v_n\{2\}$ in Pb+Pb collisions at 2.76 to 5.023 TeV are shown in the upper panel of Fig.~\ref{fig:flowratio}. The predicted increase ranges from up to 8 \% 
for $v_2$ to up to 25 \% for $v_4$. The predictions match well with the ALICE measurements for central to mid-central collisions, only in the most peripheral collisions the $\eta/s =dyn$ parametrization overestimates the data slightly, especially in the case of $v_4$, but there the experimental errors of the ratios are also quite large.

The situation is quite different in the case of Xe+Xe collisions. The ratio of the flow coefficients between the 5.44 TeV Xe+Xe and 5.023 TeV Pb+Pb collisions is shown  in the lower panel of Fig.~\ref{fig:flowratio}. The change in the flow coefficients is significantly larger than in the previous case, even if the collision energy is almost the same in Xe+Xe as in Pb+Pb collisions. The reason is that the system size is quite different when the nuclear mass number changes from $A=208$ to $A=129$. The most striking feature is the strong increase of $v_2$ in central Xe+Xe collisions compared to Pb+Pb collisions. A significant factor in the increase is the shape deformation of Xe nuclei. The deformation enhances the initial elliptic eccentricity fluctuations compared to the spherical double magic Pb nuclei. As a result the elliptic flow is 30 \% higher in the Xe case. The fact that we correctly predict this increase by taking into account the nuclear deformation is further evidence that the azimuthal asymmetries in the $p_T$ spectra are resulting from a fluid dynamical response to the initial geometry.

\subsection{Event-plane correlations, cumulants and flow-transvese-momentum correlations}
\begin{figure*}
\includegraphics[width=\textwidth]{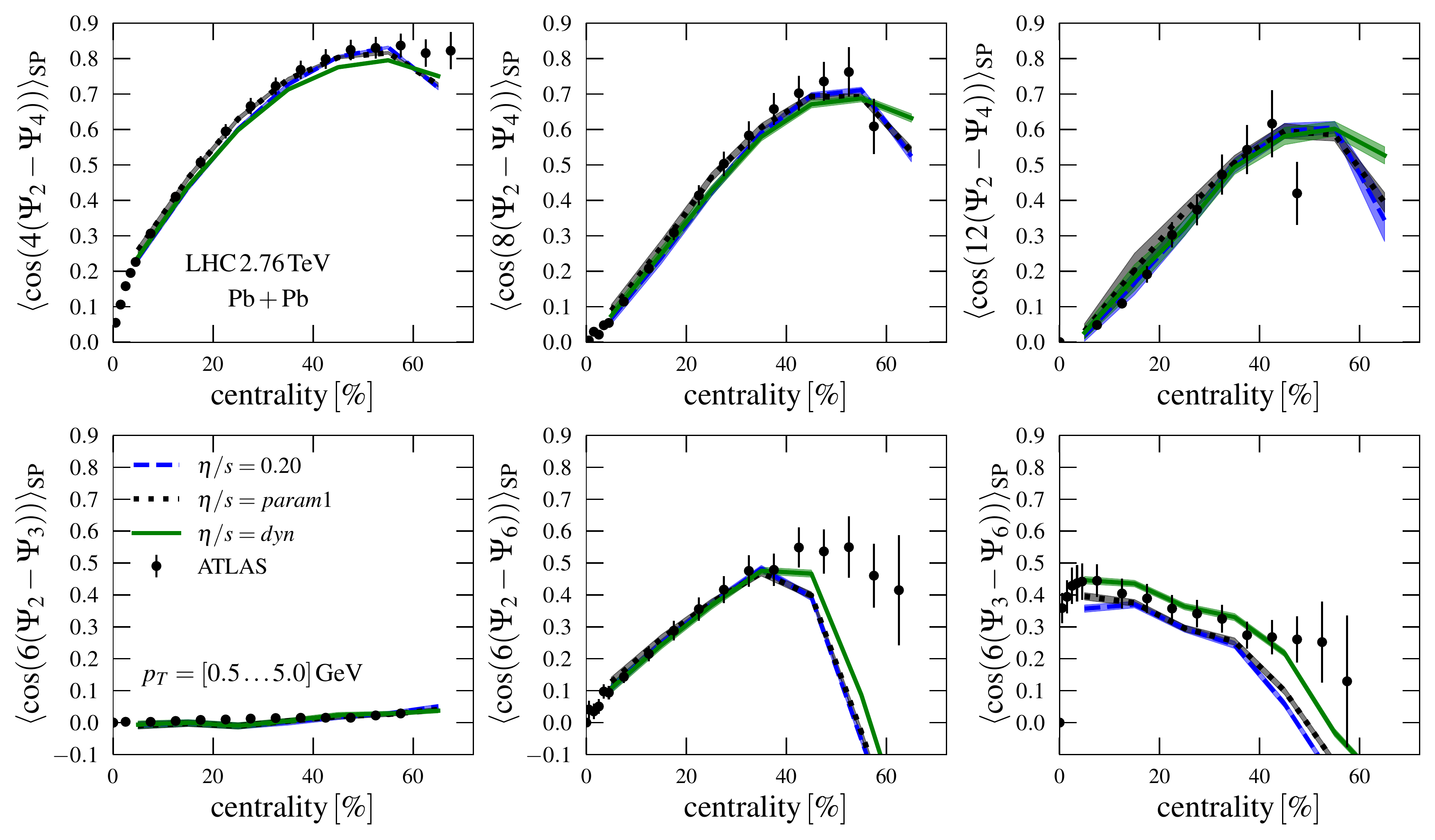}
\caption{(Color online) Event-plane correlations in 2.76 TeV Pb+Pb collisions. The data are from the
ATLAS Collaboration \cite{ATLAS:2014ndd}.}
\label{fig:EP2760}
\end{figure*}

The event-plane correlations, defined in Eq.~\eqref{eq:epcorrelation}, quantify the correlation between the event-plane angles $\Psi_n$, and also between the flow magnitudes $v_n$. The computed event-plane correlations in 2.76 TeV Pb+Pb are shown in Fig.~\ref{fig:EP2760}. Only a slight separation between the dynamical freeze-out and earlier $\eta/s(T)$ parametrizations can be seen and all parametrizations are able to describe the data. The most notable exceptions are the correlations involving the event-plane angle $\Psi_6$, which are very sensitive to $\delta f$ corrections. In these, the $\eta/s = dyn$ parametrization slightly improves the agreement with the data from the earlier works. This is mostly due the fact that the $\eta/s = dyn$ parametrization has lower shear viscosity and thus smaller $\delta f$ corrections. The event-plane correlations have only been measured for 2.76 TeV Pb+Pb collisions which is why we do not show results for other collision systems.

The symmetric cumulants, defined through Eq.~\eqref{eq:sc}, are complementary to the event-plane correlators in the sense that they depend on the correlation between the flow magnitudes $v_n$ like the event-plane correlators, but are independent of the event-plane angles. The symmetric cumulants themselves are not a measure of correlation, but depend explicitly on the magnitude of $v_n$, and not only on the degree of correlation. The corresponding correlation measure is defined through the normalized symmetric cumulants, Eq.~\eqref{eq:nsc}. 

\begin{figure*}
\includegraphics[width=\textwidth]{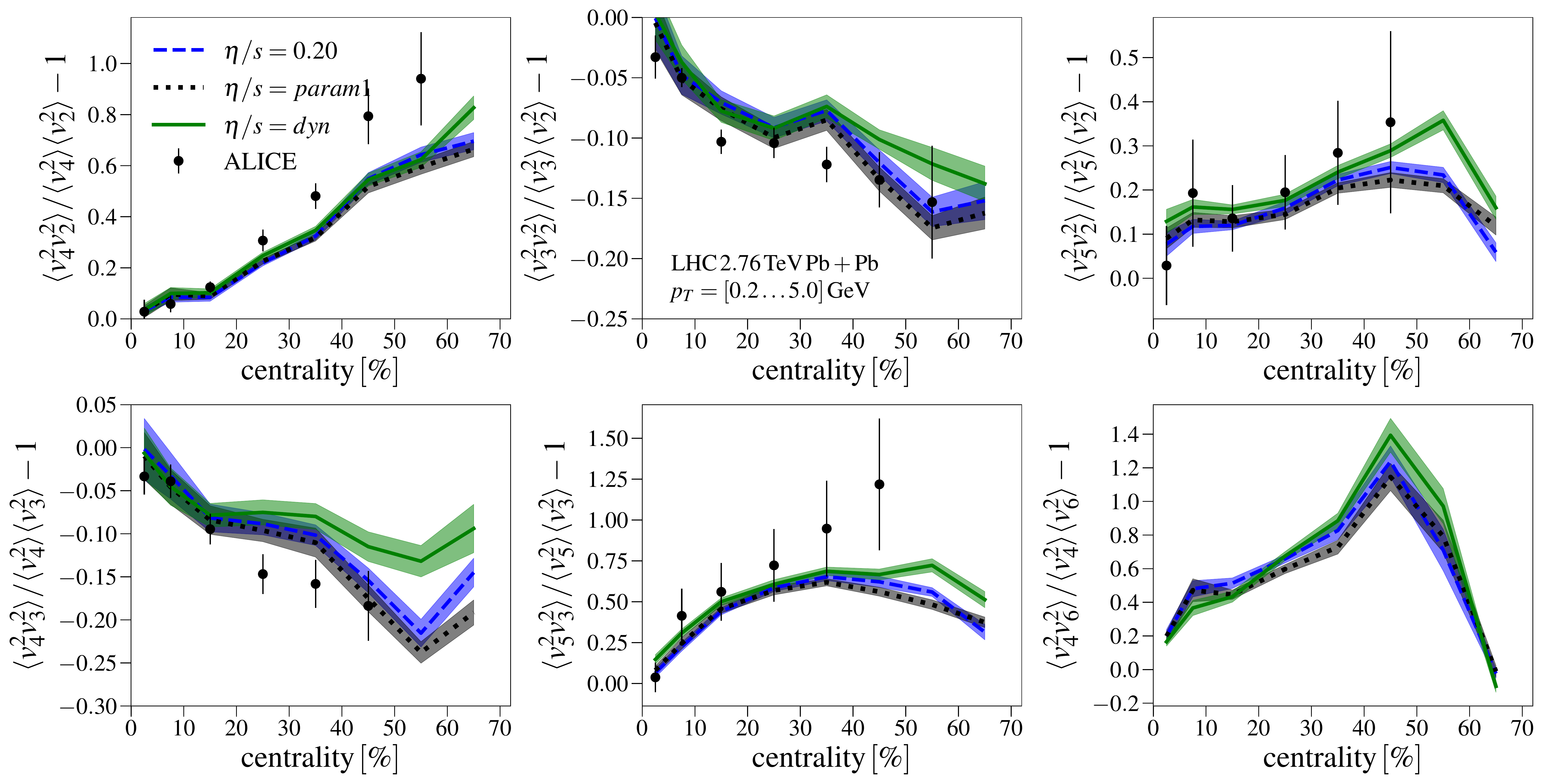}
\caption{(Color online) Normalized symmetric cumulants $NSC(n, m)$ in $2.76$ TeV Pb+Pb collisions. The data are from the ALICE Collaboration \cite{ALICE:2017kwu}.}
\label{fig:nsc2760}
\end{figure*}

\begin{figure*}
\includegraphics[width=\textwidth]{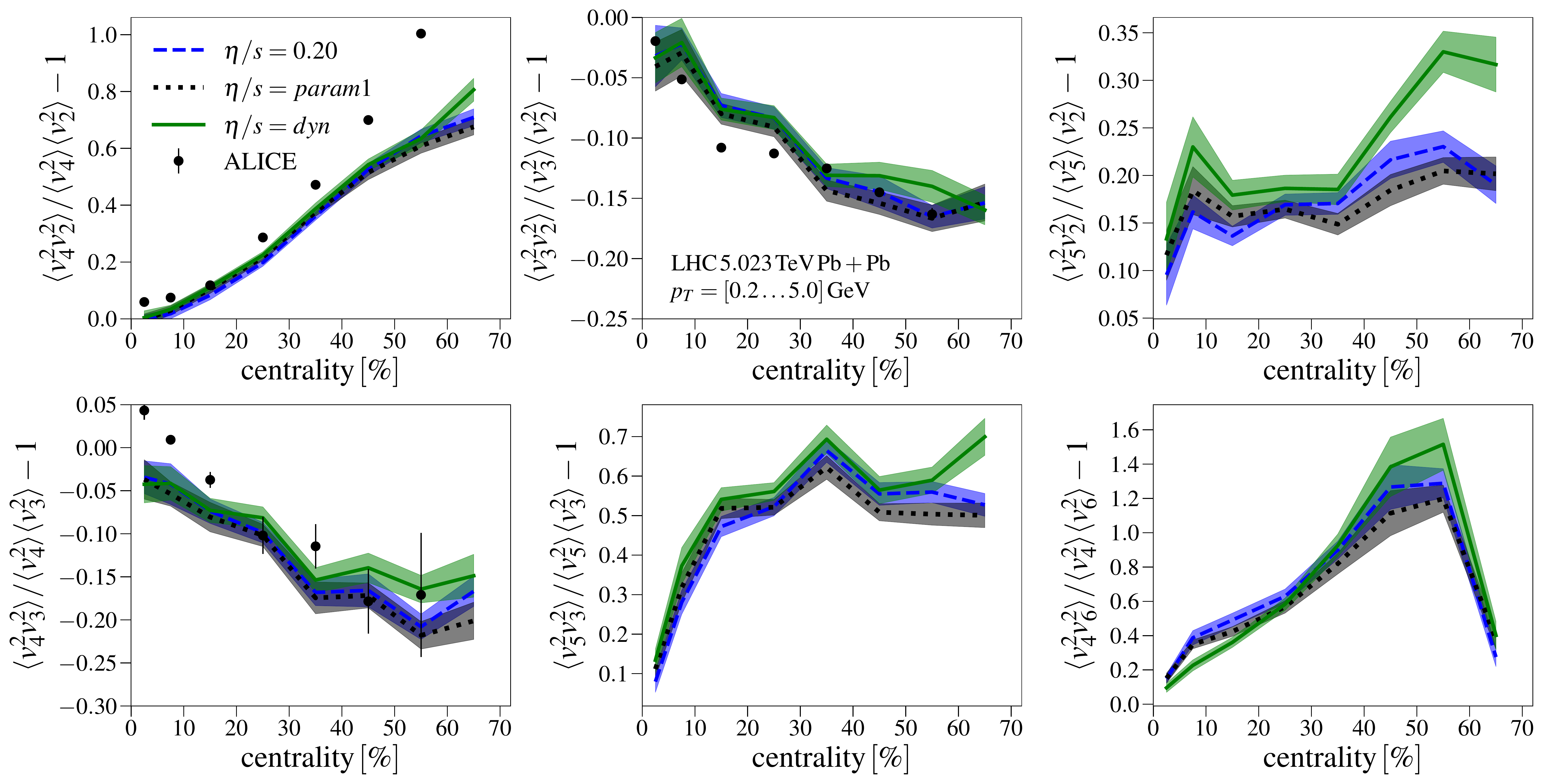}
\caption{(Color online) Normalized symmetric cumulants $NSC(n, m)$ in $5.023$ TeV Pb+Pb collisions. The data are from the ALICE Collaboration \cite{ALICE:2021adw}.}
\label{fig:nsc5023}
\end{figure*}

The normalized symmetric cumulants in 2.76 TeV Pb+Pb collisions are shown in Fig.~\ref{fig:nsc2760} compared to the ALICE data~\cite{ALICE:2017kwu}. As in the case of event-plane correlations, there are only small differences between the three $\eta/s$ parametrizations. The overall agreement between the data and the computations is good, but with a notable exception that in peripheral collisions we underpredict the $NSC(2,4)$ correlation. The collision energy dependence of the normalized symmetric cumulants is weak, as can be seen in Fig.~\ref{fig:nsc5023} where we show them in 5.023 TeV Pb+Pb collisions.

\begin{figure*}
\includegraphics[width=\textwidth]{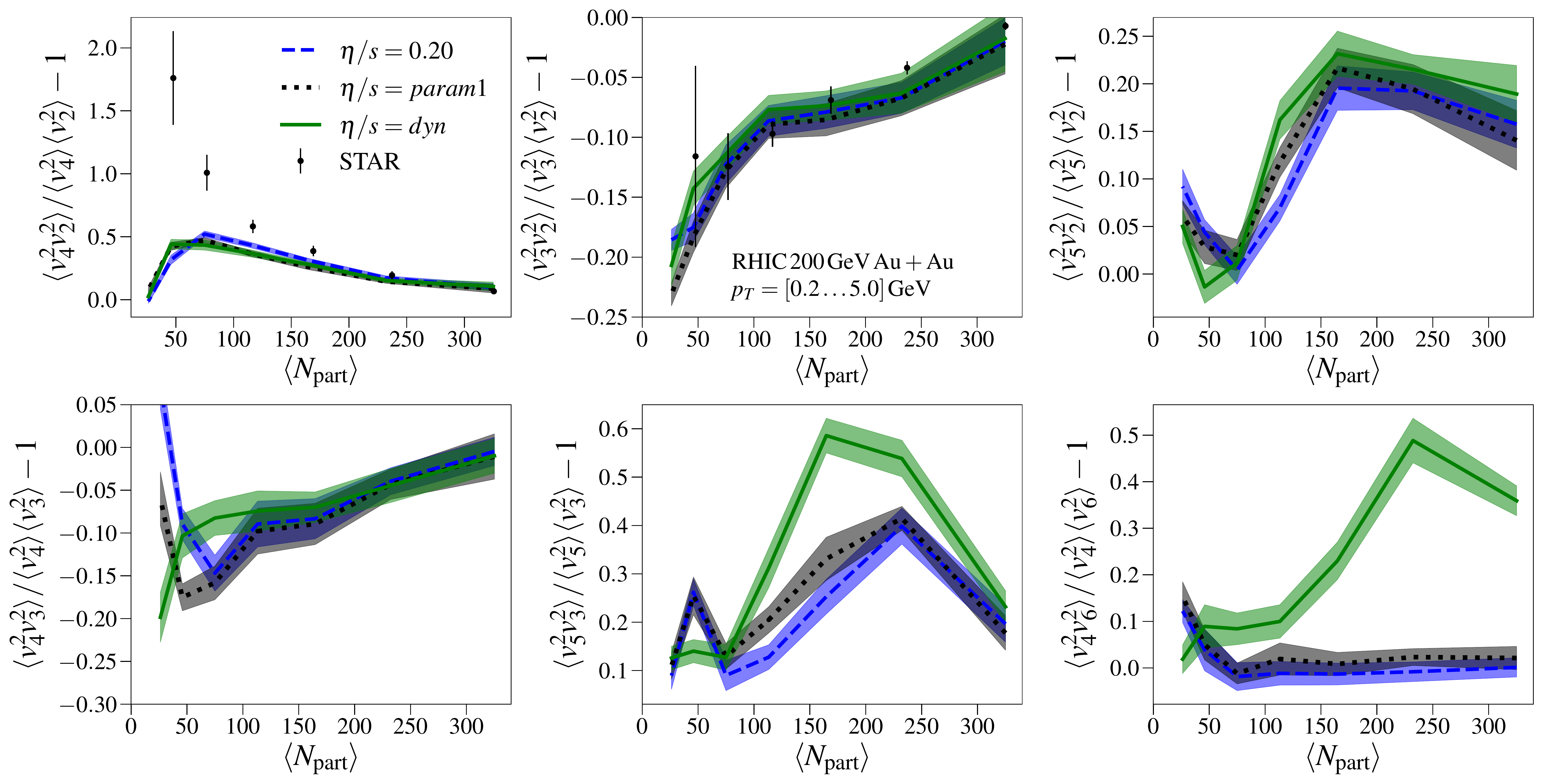}
\caption{(Color online) Normalized symmetric cumulants $NSC(n, m)$ in $200$ GeV Au+Au collisions. The data are from the STAR Collaboration \cite{STAR:2018fpo}.}
\label{fig:nscRHIC}
\end{figure*}

In Fig.~\ref{fig:nscRHIC} we show the normalized symmetric cumulants in 200 GeV Au+Au collisions. Note that here the centrality of the 
collisions is given by the number of participants, as reported by the STAR Collaboration~\cite{STAR:2018fpo}. Compared to Pb+Pb collisions we see much more 
separation between the dynamical freeze-out and earlier parametrizations for the $NSC(3,4)$, $NSC(3,5)$ and $NSC(4,6)$ correlations. The predictions for the $NSC(2,3)$ correlation is in line with the measurements while for $NSC(2,4)$ all the parametrizations clearly underestimate the data in peripheral collisions. 

\begin{figure*}
\includegraphics[width=\textwidth]{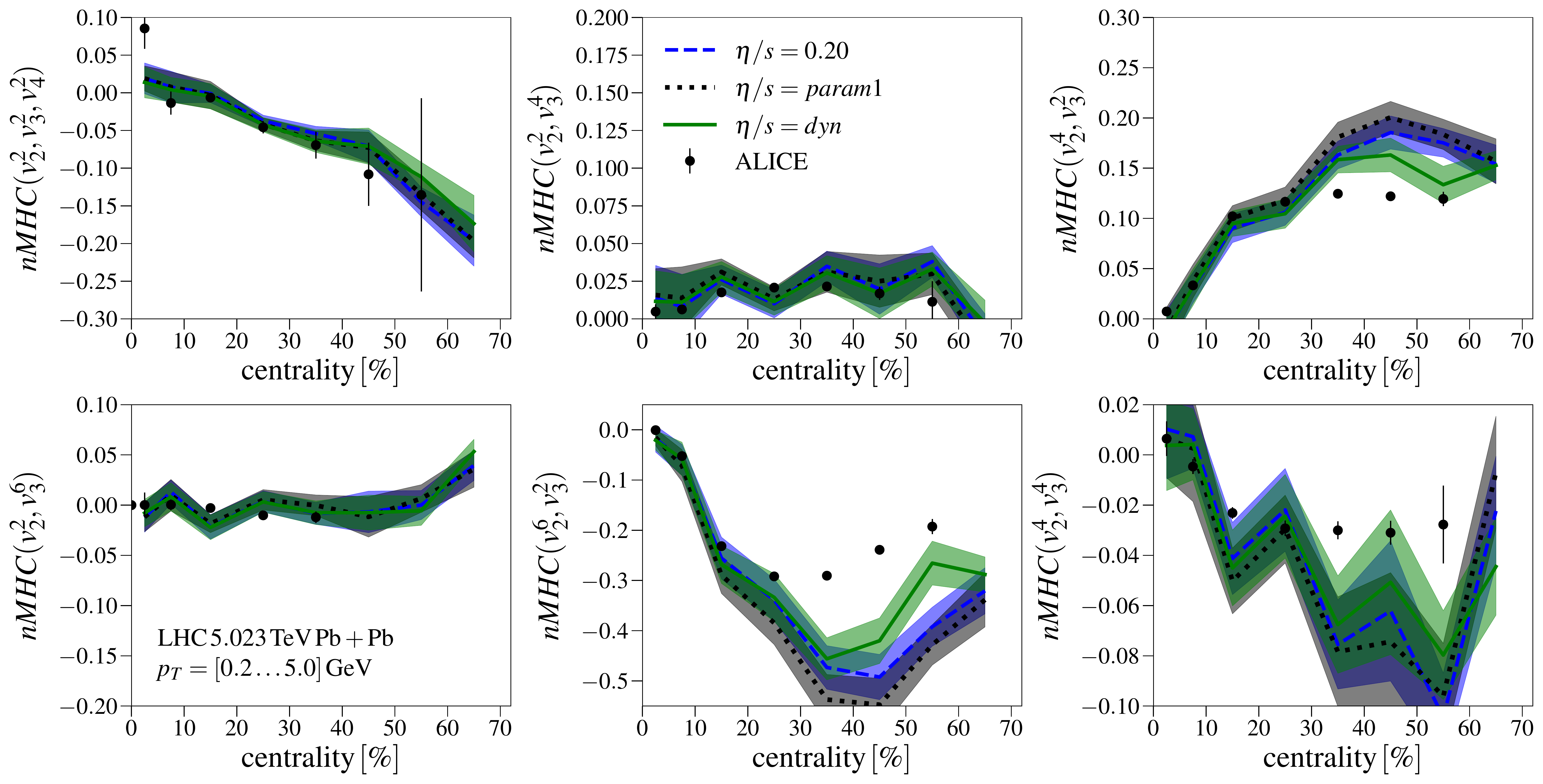}
\caption{(Color online) Normalized mixed harmonic cumulants $nMHC$ in $5.023$ TeV Pb+Pb collisions. The data are from the ALICE Collaboration \cite{ALICE:2021adw}.}
\label{fig:nmhc}
\end{figure*}

The correlations between higher order moments of two or three flow coefficients can be studied using the mixed harmonic cumulants which provide information that is independent of the normalized symmetric cumulants. The EKRT model predictions for $nMHC(v_2^2,v_3^2,v_4^2)$ and $nMHC(v_2^k,v_3^l)$ are compared against the ALICE measurements for 5.023 TeV Pb+Pb collisions in Fig.~\ref{fig:nmhc}. As can be seen there are only modest differences between the parametrizations and the statistical errors in our simulations are already quite large, especially with $nMHC(v_2^4,v_3^4)$. This is expected, since the correlations between $v_2$ and $v_3$ are thought to be more sensitive to the initial state rather than to the dynamics of the system. Our predictions seem to agree quite well with the data except for $nMHC(v_2^4,v_3^4)$, for which we predict a stronger correlation in peripheral collisions than what is measured. 

\begin{figure*}
\includegraphics[width=\textwidth]{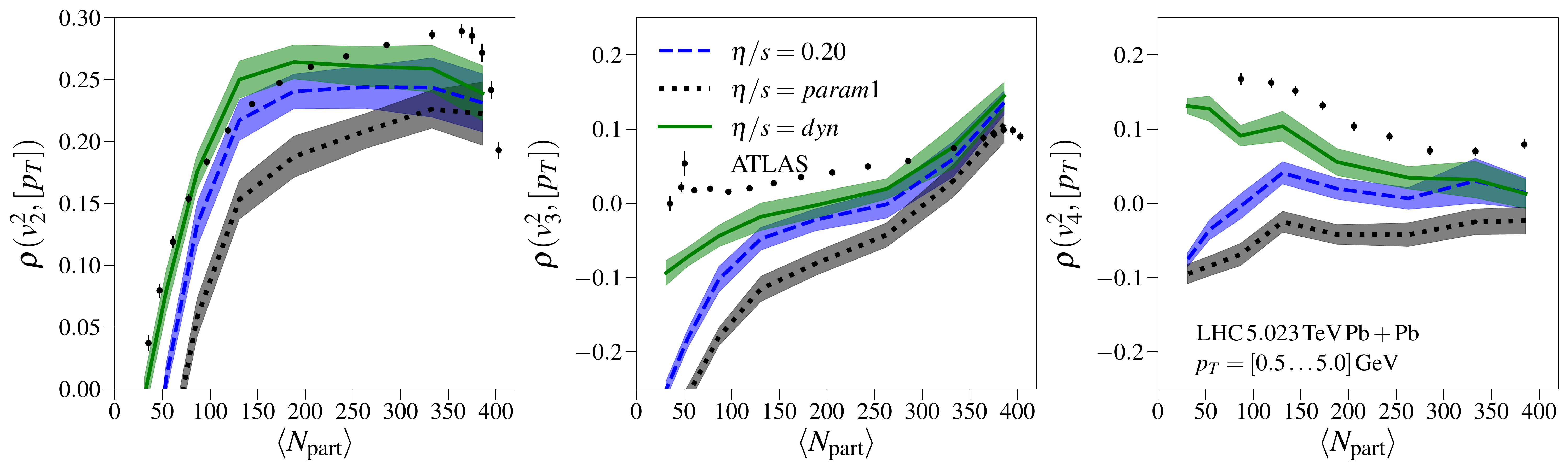}
\caption{(Color online) The flow-transverse momentum correlation coefficient $\rho(v_n\{2\}, [p_T])$ in $5.023$ TeV Pb+Pb collisions. The data are from the ATLAS Collaboration \cite{ATLAS:2019pvn}.}
\label{fig:rhov2pt}
\end{figure*}

Finally in Fig.~\ref{fig:rhov2pt} we show our predictions for the recently measured flow-transverse momentum correlations $\rho(v^2_n, [p_T])$ as a function of the number of participant nucleons in 5.023 TeV Pb+Pb collisions. These correlators describe the correlation between the average transverse momentum and the flow coefficients and thus one would expect it to be somewhat sensitive to the bulk viscosity and freeze-out criterion. The EKRT model calculations confirm this by showing an increase in all $\rho(v^2_n, [p_T])$ correlations, especially in the peripheral region. This also improves the agreement with the ATLAS measurements in peripheral collisions, even though the agreement with the data is still only qualitative. Most notably the $\eta/s = dyn$ parametrization gives the same sign as the measurements for $\rho(v^2_4, [p_T])$ in peripheral collisions.


\subsection{Higher-order flow and response coefficients}
\begin{figure*}
\includegraphics[width=\textwidth]{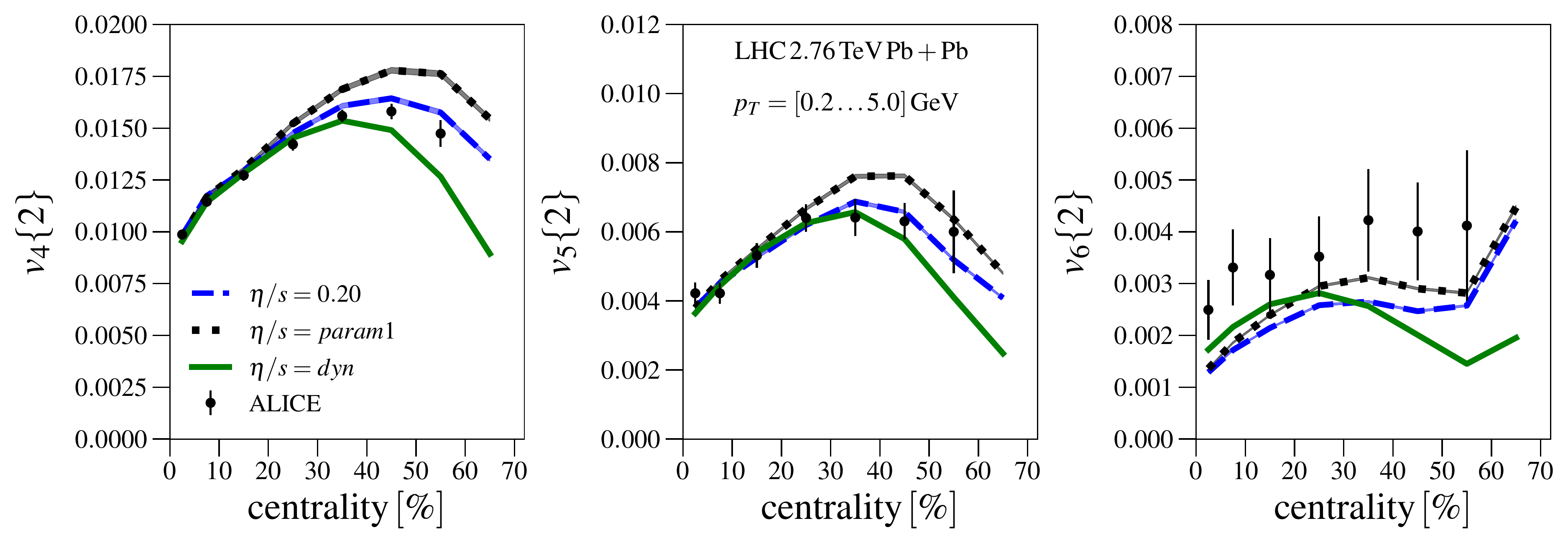}
\caption{(Color online) Higher-order flow coefficients in $2.76$ TeV Pb+Pb collisions. The data are from the ALICE collaboration \cite{ALICE:2017fcd}.}
\label{fig:nonlinearvn2760}
\end{figure*}

\begin{figure*}
\includegraphics[width=\textwidth]{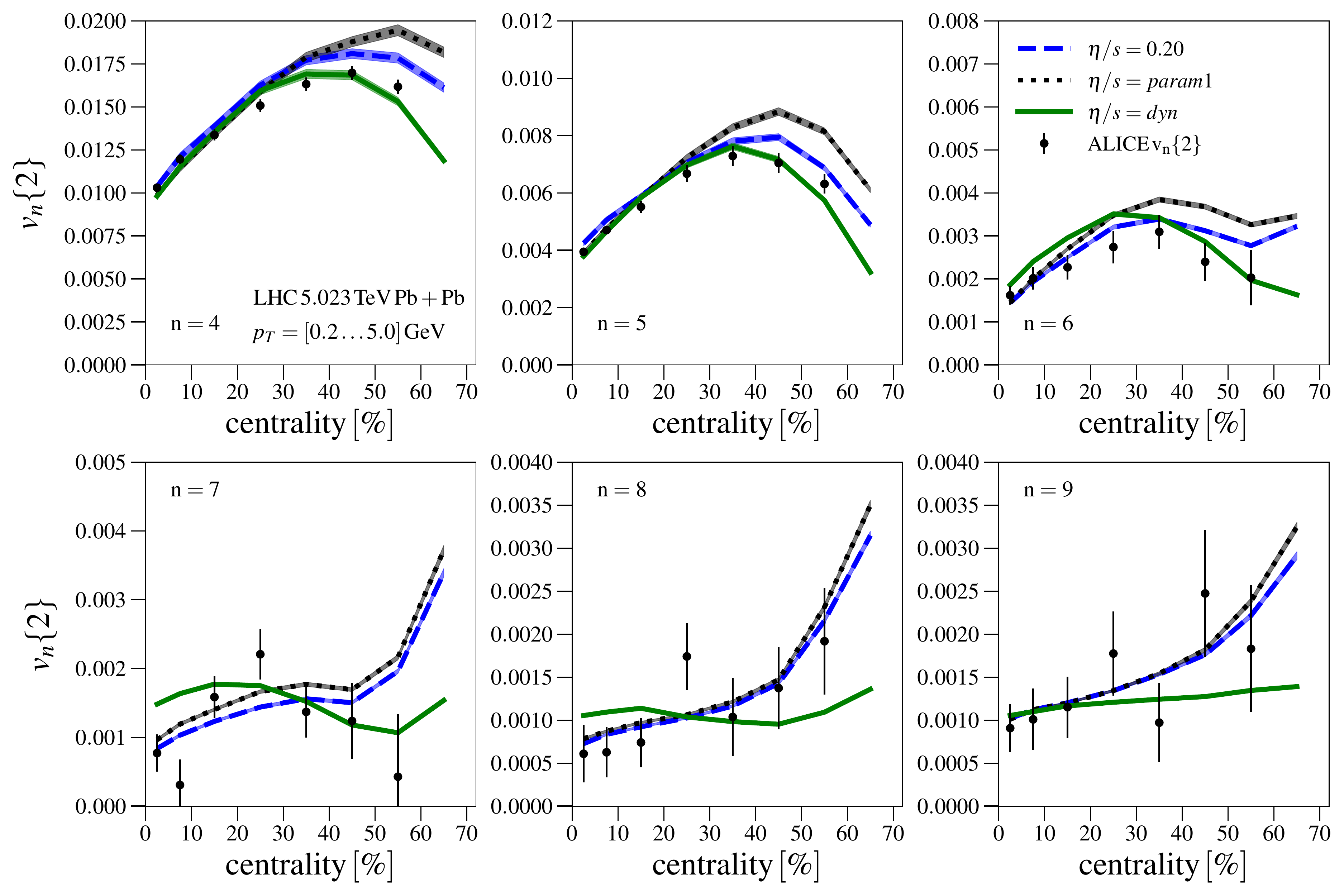}
\caption{(Color online)  Higher-order flow coefficients in $5.023$ TeV Pb+Pb collisions. The data are from the ALICE collaboration \cite{ALICE:2020sup}.}
\label{fig:nonlinearvn5023}
\end{figure*}

In Fig.~\ref{fig:nonlinearvn2760} we show the higher-order flow coefficients $v_4$, $v_5$, and $v_6$ compared to the ALICE data~\cite{ALICE:2017fcd} in 2.76 TeV Pb+Pb collisions. As can be seen in the figure, the $\eta/s=dyn$ parametrization seems to slightly underpredict the higher order flow coefficients in peripheral collisions, while the $\eta/s = 0.2$ parametrization manages to reproduce the data quite well. For $v_6$ we point out that the measured flow is larger in 2.76 TeV than in 5.023 TeV collisions, as can be seen by comparing measurements with Fig.~\ref{fig:nonlinearvn5023}, which is in conflict with the behavior of the other flow coefficients. We also note that the difference between the earlier parametrizations $\eta/s = 0.2$ and $\eta/s = param1$ is more visible here than in the case of lower-order flow coefficients.

\begin{figure*}
\includegraphics[width=0.8\textwidth]{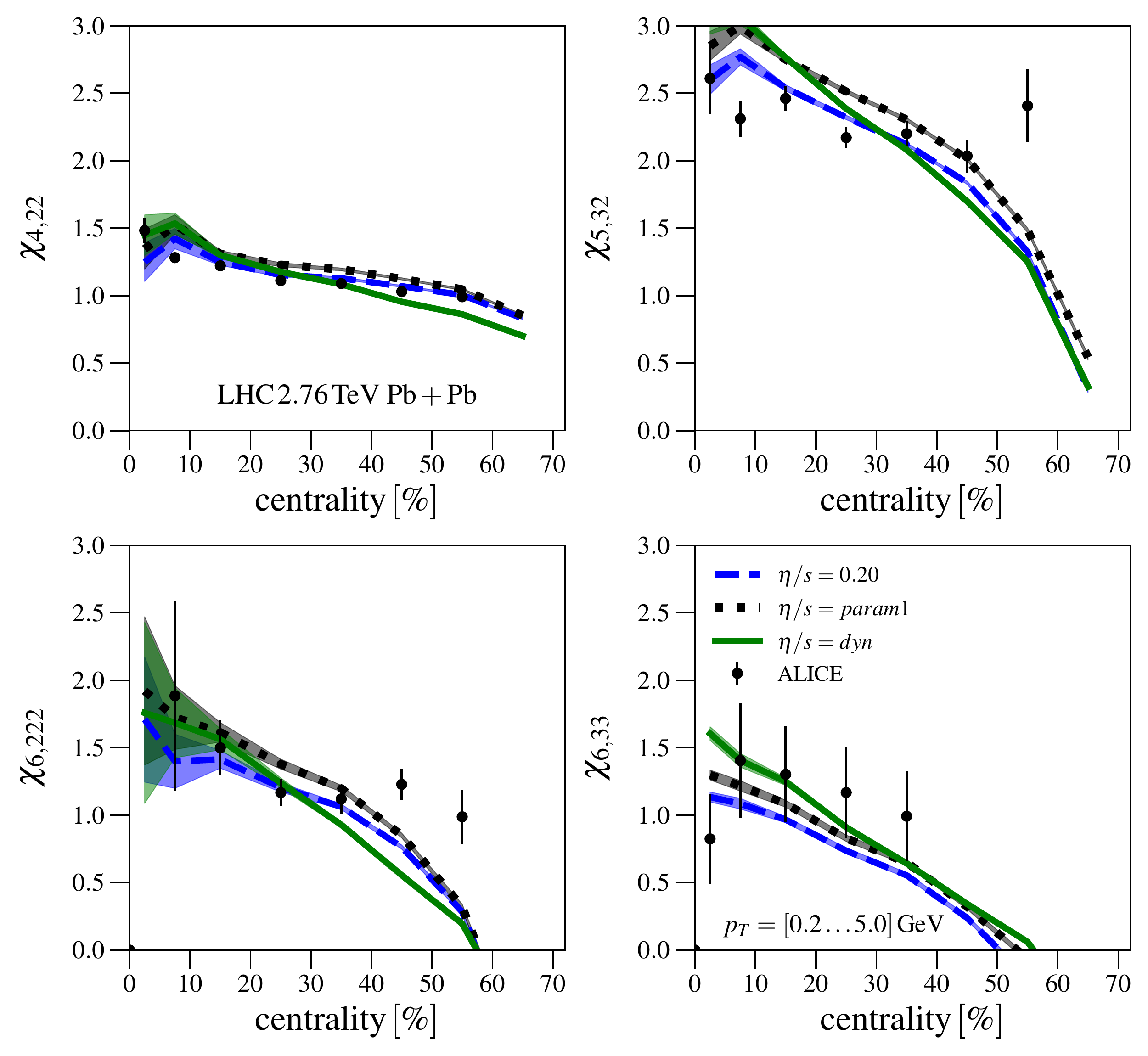}
\caption{(Color online) Non-linear flow response coefficients in $2.76$ TeV Pb+Pb collisions. The data are from the ALICE collaboration \cite{ALICE:2017fcd}.}
\label{fig:nonlinearcoeff2760}
\end{figure*}

The corresponding nonlinear response coefficients are shown in Fig.~\ref{fig:nonlinearcoeff2760}. As explained in Sec.~\ref{sec:correlators} they are closely related to the event-plane correlations, and the good agreement of the calculated response coefficients with the ALICE data is consistent with the good agreement between the calculated and the measured ATLAS event-plane correlations in Fig.~\ref{fig:EP2760}.  


\begin{figure*}
\includegraphics[width=0.8\textwidth]{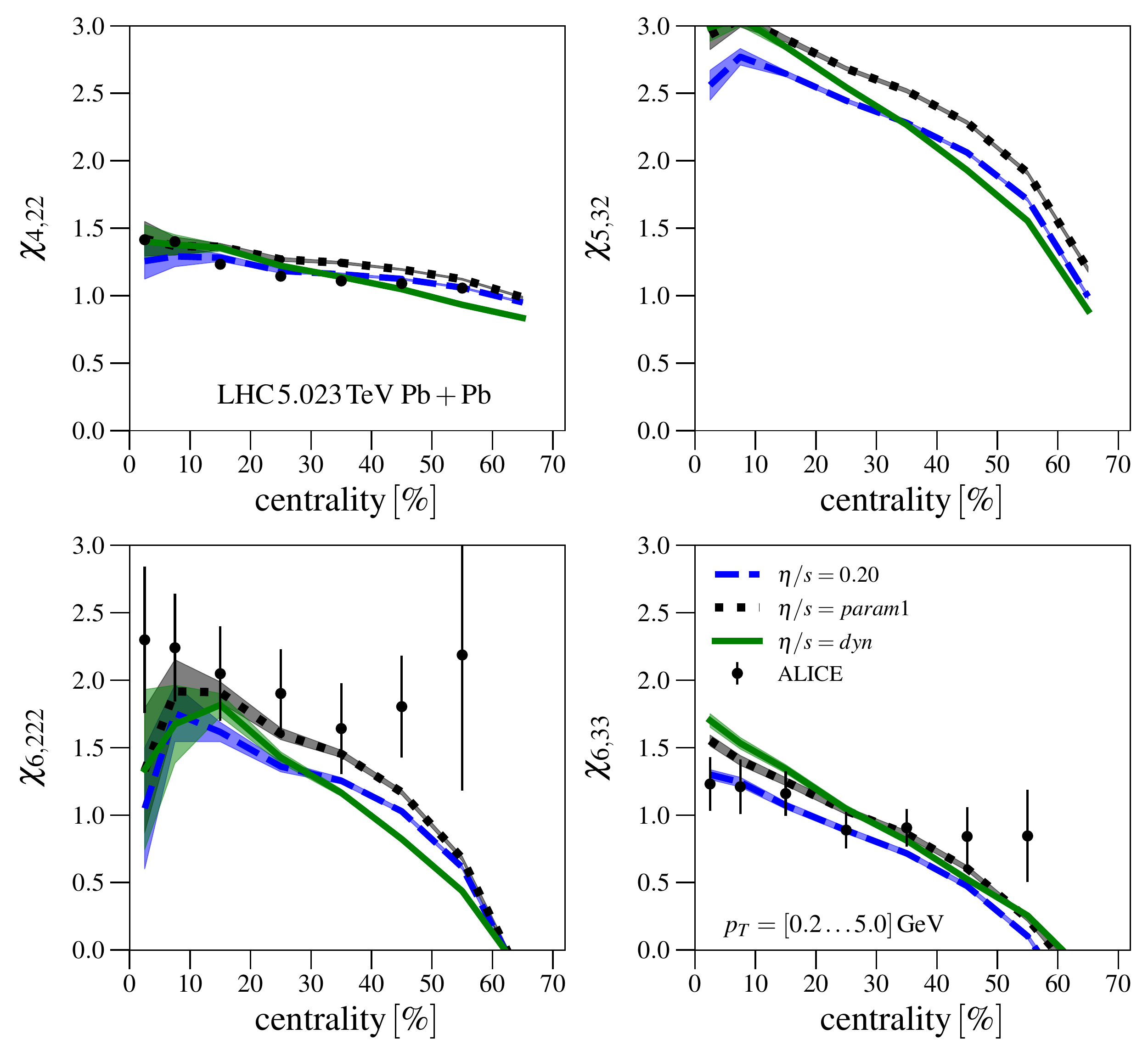}
\caption{(Color online) Non-linear flow response coefficients in $5.023$ TeV Pb+Pb collisions. The data are from the ALICE collaboration \cite{ALICE:2020sup}.}
\label{fig:nonlinearcoeffs5023}
\end{figure*}

The same flow and response coefficients as above, but for 5.023 TeV Pb+Pb collisions, are shown in Figs.~\ref{fig:nonlinearvn5023} and \ref{fig:nonlinearcoeffs5023}, respectively. Together with other higher order flow harmonics we also show $v_7$, $v_8$, and $v_9$, which are only measured for the 5.023 TeV energy. Here we see that the parametrization that uses dynamical freeze-out predicts the higher order flow coefficients quite well while the parametrizations from earlier works are slightly above the measurements. 

The response coefficients are not directly proportional to the magnitude of the flow coefficients, or the proportionality is partly canceled by the normalization. That is to say that the agreement in the response coefficients with the ALICE data is similar as at the lower collision energy even though we cannot exactly reproduce the higher order $v_n$'s for both collision energies simultaneously. 

The overall agreement with the higher-order flow coefficients with the data is quite similar for both the earlier and current EKRT setup. The improvements due to the dynamical decoupling are not as clear as for $v_2$. However, the differences between the parametrizations are also larger, highlighting the fact that higher-order coefficients, and their $\sqrt{s_{\rm NN}}$ dependence give important constraints to the determination of shear viscosity.


\section{Summary and Conclusions}
\label{sec:conclusions}

We have presented the results for the low-$p_T$ observables in Pb+Pb, Au+Au, and Xe+Xe collisions at RHIC and LHC energies from the fluid dynamical computations using the NLO pQCD based EKRT model for the initial conditions. Compared to the previous EKRT works in Refs.~\cite{Niemi:2015qia, Niemi:2015voa, Eskola:2017bup} we have now added the bulk viscosity together with the dynamical decoupling conditions to improve the validity of our model in peripheral collisions. 

The overall agreement of the computed results with the data is very good in particular for the $\sqrt{s_{\rm NN}}$, $A$, and centrality dependence of the charged hadron multiplicity. This is mainly a feature of the EKRT initial conditions. The main uncertainty in the EKRT model is the $K_{\rm sat}$ parameter in the saturation condition, but this can be essentially fixed from one measurement of charged hadron multiplicity. Even if the value of $K_{\rm sat}$ depends on the $\eta/s$ parametrization through the entropy production during the fluid dynamical evolution, the final results for the $\sqrt{s_{\rm NN}}$, $A$, and centrality dependence are practically independent of the $K_{\rm sat}$ value, making them very robust predictions of the EKRT model.

The most significant effect of the dynamical freeze-out can be seen in the absolute magnitude of the flow coefficients $v_n$. We have demonstrated that we can reproduce the experimental data for $v_2$ and $v_3$ across the centrality range $0-80$ \% in all the collision systems with the exception of peripheral RHIC collisions. This is a significant improvement from the constant-temperature freeze-out which only manages to describe the data up to the $30-40$ \% centrality class. The higher harmonics $v_4$, $v_5$, and $v_6$ are quite similarly described by both the earlier computations and the current setup, but the differences between the $\eta/s$ parametrizations are also more pronounced. On the other hand, the relative increase of the flow coefficients from 2.76 TeV Pb+Pb to 5.023 TeV Pb+Pb and 5.44 TeV Xe+Xe collisions is well described in all the centrality classes shown here. The addition of the dynamical freeze-out together with the bulk viscosity has also made it possible to improve the simultaneous agreement of the identified particle multiplicities and the mean transverse momenta with the measurements.

We have also shown the EKRT model predictions for the most recent correlation measurements. Our results for the symmetric cumulants, the mixed harmonic cumulants, the response coefficients, and closely related event-plane correlators are very similar to the earlier EKRT results and the agreement with the data remains reasonably good. The most notable differences are in $NSC(2,4)$ correlators in peripheral collisions, where the predictions are visibly below the experimental data. The effect of the dynamical freeze-out and the bulk viscosity can be seen in the flow-transverse-momentum correlators $\rho(v_n^2, [p_T])$, where we demonstrated a better quantitative agreement with the experimental measurements in peripheral collisions than given by the previous EKRT computations. Especially, we obtained the correct sign in $\rho(v_4^2, [p_T])$ correlation in peripheral collisions.

In conclusion, we have introduced dynamical freeze-out conditions to model the decoupling of the fluid to free hadrons. In particular, the aim was to capture the essential features of the decoupling that take into account the system size variations at different collision energies and centralities. The clear benefit here is that it allows us to keep the transport coefficients continuous throughout the whole temperature range, without unphysical discontinuities that can appear at a switching between fluid dynamics and hadron cascade. At the same time it is then possible to use the measured data to constrain the QCD matter transport properties also in the hadronic phase.

We emphasize that in spite of the extensive iteration work done, the parametrizations shown here do not necessarily represent the absolute best fit to the data. For that we would need to do a full statistical global Bayesian analysis of the parameter space. This we have left as a future work. However, we have demonstrated that we can reproduce the measured LHC and RHIC low-$p_T$ observables reasonably well, and the dynamical decoupling leads to quite a different spacetime picture compared to many hydro+cascade models. Instead of a very viscous hadronic evolution directly after the low-viscosity QGP evolution, in the picture presented here the low-viscosity evolution can extend to quite low temperatures on the hadronic side. 

\acknowledgments
We thank Pasi Huovinen for discussions, and for providing us with the EoS tables. We acknowledge the financial support from the Jenny and Antti Wihuri Foundation, and the Academy of Finland Project No. 330448 (K.J.E.). This research was funded as a part of the Center of Excellence in Quark Matter of the Academy of Finland (Project No. 346325). This research is part of the European Research Council Project No. ERC-2018-ADG-835105 YoctoLHC. The Finnish IT Center for Science (CSC) is acknowledged for computing time through Project jyy2580.

\end{document}